\documentclass[aps,showpacs]{revtex4}
\usepackage{dcolumn}
\usepackage{multirow}
\usepackage{bm}
\usepackage[dvips]{graphics}
\usepackage{amssymb}
\usepackage{extarrows}
\begin{document}
\title{Ground states of a mixture of two species of spin-$1$ Bose gases with interspecies spin exchange in a magnetic field}
\author{Yu Shi}
\email{yushi@fudan.edu.cn}
\affiliation{
Department of Physics, Fudan University, Shanghai 200433, China}
\author{Li Ge}
\affiliation{
Department of Physics, Fudan University, Shanghai 200433, China}
\begin{abstract}

We consider a mixture of two species of spin-1 atoms with both interspecies and intraspecies spin exchanges in a weak magnetic field. Under the usual single mode approximation, it can be reduced to a model of coupled giant spins. We find most of its ground states.
This is a complicated problem of energy minimization, with three quantum variables under constraints, i.e. the total spin of each species and the total spin of the whole mixture, as well as four parameters, including intraspecies and  interspecies spin coupling strengths and the magnetic field. The quantum phase diagram is very rich. Compared with the case without a magnetic field, the ground states are modified by a magnetic field, which also  modifies the ground state boundaries or introduces new crossover regimes on the phase diagram.
Without interspecies spin coupling,  the quantum phase transitions existing in absence of a magnetic field disappear when a magnetic field is applied, which leads to crossover regimes in the phase diagram.
Under ferromagnetic interspecies spin coupling, the ground states remain disentangled no matter whether there is a magnetic field.
For antiferromagnetic interspecies spin coupling, a magnetic field entangles the ground states in some parameter regimes.
When the intraspecies spin couplings are both ferromagnetic, the quantum phase transition between antiferromagnetic and zero  interspecies spin couplings survives the magnetic field. When the intraspecies spin couplings are both antiferromagnetic,
a magnetic field induces new quantum phase transitions between antiferromagnetic and zero  interspecies spin couplings.

\end{abstract}


\maketitle

\section{Introduction}

Spinor Bose gases have been extensively studied since a decade ago when it was discovered that they display remarkable spin correlations because of spin-exchange scattering between atoms~\cite{ho1,law,koashi,ho2,hoyin,zhou,ueda2,yang,spinor,pethick}. However, there have not been many investigations on many-body phenomena in mixtures of different spinor Bose gases with interspecies spin exchanges. In the first instance,  spin-exchange scattering between distinguishable atoms have been less studied, perhaps because of incomplete information on inter-atomic potential.  To motivate more interests in this direction, we note that interspecies spin-exchange interaction can be significant. The previous experiments on multi-component Bose gases often had atom loss due to spin exchanges~\cite{spinor,myatt}.
There were early calculations indicating that the  cross-sections  of spin-exchange scattering  between different atoms may not be smaller than those between identical atoms~\cite{dalgarno}. Recently, there were more calculations as mentioned in the following, though motivated by studying a mixture of two species of atoms in frozen spin states, for which spin-exchange scattering is regarded as inelastic. A calculation for $^{23}$Na-$^{85}$Rb scattering indicated a  quite large difference between  scattering lengths of electronic singlet and triplet states~\cite{weiss}. This paper also reported a small value of such a singlet-triplet difference of scattering lengths for $^{23}$Na-$^{87}$Rb scattering, but contrary result was later reported~\cite{pashov}.  Another calculation  found  significant singlet-triplet differences of scattering lengths  for X-$^{133}$Cs scattering, where X=$^6$Li, $^7$Li, $^{39}$K, $^{41}$K, $^{85}$Rb and $^{87}$Rb~\cite{zanelatto}.  Experimentally, significant differences between singlet and triplet scattering lengths have been observed in $^{41}$K-$^{87}$Rb, $^{40}K$-$^{87}$Rb and $^6$Li($^7$Li)-$^{23}$Na mixtures~\cite{ferrari,simoni,inouye,gacesa}, implying significant interspecies spin exchanges. Spin-changing scattering was also observed in $^7$Li-$^{133}$Cs \cite{mudrich}.
Moreover, heteronuclear Feshbach resonances can be implemented~\cite{chin}, which can enhance both elastic and inelastic collision rates~\cite{li,marzok,deh}. To our understanding, some recent experimental set-ups on multi-species Bose gases have come to close to what we need to  realize a mixture of two species of spinor gases with interspecies spin exchange~\cite{recentexp}.

Further researches on mixtures of spinor gases with interspecies spin exchanges can be motivated by novel many-body quantum phenomena in such mixtures, as demonstrated first in a model of a mixture of pseudospin-$\frac{1}{2}$ atomic gases, where interspecies spin exchange leads to richer ground states and phenomena, especially Bose-Einstein condensation (BEC) with interspecies quantum entanglement, which was dubbed entangled Bose-Einstein condensation (EBEC)~\cite{shi0,shi1,shi2,shi4}. As usual, two subsystems constituting the total system are entangled if the state of the total system is not a direct product of those of the subsystems. Otherwise, they are called disentangled.

This line of researches has been extended to a mixture of two species of spin-$1$ atomic gases~\cite{shi4a,shi5}, in which the interspecies spin coupling is simply of Heisenberg form~\cite{luo}. In the usual approach of single orbit-mode approximation, most of the exact ground states in absence of a magnetic field have been found~\cite{shi4a,shi5}. However, in presence of a magnetic field, only two special parameter regimes have been considered~\cite{shi4a}. Given that the magnetic field effect is an important issue,
in the present paper, we systematically study the ground states in presence of a weak  magnetic field, and find out how a magnetic field affects the ground states and phase diagrams of a spinor mixture with interspecies spin exchanges.

The rest of the paper is organized as the following.  To make the paper self-contained, we set the stage in Sec.~\ref{introh}. Then we discuss in Sec.~\ref{no} the ground states of a mixture of two spin-$1$ Bose gases in a magnetic field, but without interspecies spin coupling. In Sec.~\ref{ferro}, we find the ground states of a mixture with  ferromagnetic interspecies spin coupling in a magnetic field, based on the calculations detailed in the Appendix A. For antiferromagnetic interspecies spin coupling $c^{ab}>0$, we divide the range of  $c^{ab}$ to three intervals.  In Sec.~\ref{anti1}, based on the calculations detailed in the Appendices B, C, D and E,  we find the ground states of a mixture with  $0 < c^{ab} \leq 2\gamma B$, where $\gamma$ is the gyromagnetic ratio, $B$ is the magnitude of the field. In Sec.~\ref{anti2}, we make some brief discussions on the regime $c^{ab} > 2\gamma B$.  In Sec.~\ref{qpt},  quantum phase transitions are described. The issue of characterizing interspecies entanglement is discussed in Sec. \ref{entanglement}.   A summary is made in Sec.~\ref{summary}.

\section{The System  \label{introh}}

Consider a mixture of two species $a$ and $b$ of  spin-1 atoms,  whose numbers
$N_a$ and $N_b$ are conserved respectively. The single-atom Hamiltonian of species  $\alpha$ ($\alpha = a, b$) is
\begin{equation}
h_{\alpha}=-\frac{\hbar^2}{2m_{\alpha}}\nabla^2+U_{\alpha} (\mathbf{r})
-\gamma_{\alpha}\mathbf{B} \cdot
\mathbf{F}_{\alpha},
\end{equation}
where $\mathbf{B}$ is a
uniform magnetic field,  $m_{\alpha}$, $\gamma_{\alpha}>0$, $U_{\alpha}$ and $\mathbf{F}_{\alpha}$ are the  mass, the
gyromagnetic ratio, the external potential and the single-spin operator,
respectively,  for an atom of species $\alpha$.  With $\psi_{\alpha\mu}$ representing the field operator corresponding
to spin $\mu$ component of species $\alpha$ ($\mu = -1, 0, 1$), the many-body Hamiltonian is
\begin{equation}
{\cal H}
= \sum_{\alpha=a,b} {\cal H}_\alpha + {\cal H}_{ab},
\end{equation}
where
\begin{equation}
{\cal H}_{\alpha} = \int d\mathbf{r}
\psi^{\dagger}_{\alpha\mu}h_{\alpha} (\mathbf{r})_{\mu\nu}
\psi_{\alpha\nu} + \frac{1}{2} \int d\mathbf{r}
\psi^{\dagger}_{\alpha\mu}\psi^{\dagger}_{\alpha\rho}
(\bar{c}_0^{\alpha}\delta_{\mu\nu}
\delta_{\rho\sigma}+\bar{c}_2^{\alpha}
\mathbf{F}_{\alpha\mu\nu}\cdot\mathbf{F}_{\alpha\rho\sigma})
\psi_{\alpha\sigma}\psi_{\alpha\nu}
\end{equation}
is the usual Hamiltonian of spin-1 atoms~\cite{ho1},
\begin{equation}
{\cal H}_{ab}= \int d\mathbf{r}
\psi^{\dagger}_{a\mu}\psi^{\dagger}_{b\rho}
(\bar{c}_0^{ab}\delta_{\mu\nu} \delta_{\rho\sigma}+\bar{c}_2^{ab}
\mathbf{F}_{a\mu\nu}\cdot\mathbf{F}_{b\rho\sigma})
\psi_{b\sigma}\psi_{a\nu}
\end{equation}
is the interspecies interaction~\cite{shi4a}, where $\bar{c}^\alpha_0$ and $\bar{c}^\alpha_2$ are expansion coefficients in terms of powers of dot product of the single-spin matrices of  two  atoms of species $\alpha$ and are linear combinations of singlet and triplet scattering lengths,  $\bar{c}^\alpha_2$ is  proportional to the differences between triplet and single scattering lengths of intraspecies scattering~\cite{ho1,pethick}, $\bar{c}^{ab}_0$ and $\bar{c}^{ab}_2$ are similar quantities for scattering between an  $a$-atom and a $b$-atom, $\bar{c}^{ab}_2$ is  proportional to the differences between triplet and single scattering lengths of interspecies scattering,  and it has been shown that the coefficient of  $(\mathbf{F}_{a}\cdot\mathbf{F}_{b})^2$ is zero~\cite{luo}.

For each species and each spin state, we follow the usual single mode approximation for the single-particle orbital wave function,  and the usual assumption that this single particle orbital wave function is independent of spin. Therefore we have
$\psi_{\alpha\mu}(\mathbf{r})=
\alpha_{\mu}\phi_{\alpha}(\mathbf{r})$, where $\alpha_{\mu}=a_{\mu},
b_{\mu}$ is the annihilation operator and  $\phi_{\alpha}$ is the
lowest single-particle orbital wave function for species $\alpha$ and spin-independent. Then the Hamiltonian can be simplified as
\begin{equation}
{\cal H} = \frac{c^{a}}{2} \mathbf{S}_a^2 +
\frac{c^{b}}{2}\mathbf{S}_b^2 + c^{ab} \mathbf{S}_a \cdot
\mathbf{S}_b - \gamma \mathbf{B}\cdot \mathbf{S}_{a} - \gamma
\mathbf{B}\cdot \mathbf{S}_{b}, \label{spinham}
\end{equation}
where a constant is neglected,
\begin{equation}
\mathbf{S}_{\alpha}=
\alpha^{\dagger}_{\mu}\mathbf{F}_{\mu\nu}\alpha_{\nu}
\end{equation}
is the total spin operator for species  $\alpha$,
$c^{\alpha}=\bar{c}_2^{\alpha}\int
d^3r |\phi_{\alpha}|^4$ is the intraspecies spin coupling strength, $c^{ab}=\bar{c}_2^{ab}\int d^3 r
|\phi_{a}\phi_b|^2$ is the interspecies spin coupling strength,  and we have set $\gamma_a=\gamma_b=\gamma$, as indeed so for atoms with a same nuclear spin. Here we have neglected the quadratic Zeeman effect. This is reasonable under certain circumstances, as can be estimated by using parameter values for Na~\cite{spinor}.
The quadratic Zeeman energy is $\hat{q}B_0^2$, where $\hat{q}=278 Hz/G^2$, $B_0$ can be $10$mG to $500$mG, hence the quadratic Zeeman energy is about $2.78\times 10^{-2}$ to $70$Hz.
The linear Zeeman energy is about $1$ to $100$HZ.  Therefore it is easy to reach the regime where the quadratic Zeeman effect is negligible.

$S_a$, $S_b$ together with the total spin $S$
and its $z$-component $S_z$ are all good quantum numbers, as $\mathbf{S}_a^2$, $\mathbf{S}_b^2$ and $\mathbf{S}^2$ all commute with the Hamiltonian (\ref{spinham}). However it should be noted that $S_a$ and $S_b$ are not fixed numbers, as in the case of pseudospin-$\frac{1}{2}$ atoms, for which one can find $S_a=N_a/2$ and $S_b=N_b/2$. In the present case, $S_a$, $S_b$ and $S$ should all be determined by minimizing the energy.

In the presence of a magnetic field,
for a
given $S$, $S_z=S$ minimizes the energy. With $S_a$, $S_b$, $S$ and $S_z$ all being good quantum numbers, the ground state is
\begin{equation}
|G\rangle = |S_a^m,S_b^m,S^m,S^m\rangle, \label{gs}
\end{equation}
where ${\cal S}_a^m$, ${\cal
S}_b^m$ and ${\cal S}^m$ are, respectively, the values of $S_a$,
$S_b$ and $S$ that minimize the energy
\begin{equation}
\begin{array}{c}
\displaystyle
E=\frac{c^{a}-c^{ab}}{2 }
S_a(S_a+1)+ \frac{c^{b}-c^{ab}}{2 }S_b(S_b+1)
\\
\displaystyle
+ \frac{c^{ab}}{2} S(S+1) - \gamma B S,
\end{array} \label{e}
\end{equation}
under the constraints
\begin{equation}
|S_a-S_b|  \leq S \leq S_a+S_b. \label{srange}
\end{equation}

Note that the existence of  three quantities $S_a$, $S_b$ and $S$ with the constraint (\ref{srange}) as well as the limited ranges  of $S_a$ and $S_b$, and the dependence on the three parameters $c^a$, $c^b$ and $c^{ab}\neq 0$  makes this minimization problem highly nontrivial. We have managed to solve this problem in most of the parameter regimes, as reported in  in Appendices. Before discussing these cases of $c^{ab}\neq 0$, we shall first take a look at the case of $c^{ab}= 0$.

As we shall discuss different ground states in different regimes  of the parameter space, some explanation of the nomenclature is in order here. The ground states in two neighboring parameter regimes are said to be continuously connected if each of them approaches the ground state on the boundary, when the parameters approach the boundary. It is then said that they belong to a same quantum phase.  In contrast, if the two ground states in the two neighboring regimes approach different limits when the parameters approach the boundary, it is said that there is a discontinuity or quantum phase transition.  There are several cases of discontinuity, for example, the two limits may be both different from the that on boundary, and they may also be two of the degenerate ground states on the boundary, besides, there is also the case that the ground state in one of the regimes approaches  a  ground state  on the boundary, while the ground state in the other regime approaches a different limit.

The most interesting ground states in our system are those of EBEC, i.e. BEC with interspecies entanglement. Note that throughout this paper,  a state which may be entangled is written in the the general form, i.e. $|S^m_a,S^m_b,S^m,S^m\rangle$. A state which is certainly disentangled is written in the form of  $|S_a^m,S_a^m\rangle_a|S_b^m,S_b^m\rangle_b$

\section{$c^{ab}=0$ \label{no} }

Without spin-exchange interaction between the two species, i.e. $c^{ab}=0$,  the two species can be considered independently. The ground states are all disentangled.  We have $E = E_a + E_b$, with
\begin{equation}
\begin{array}{rl}
\displaystyle
E_{\alpha} & = \frac{c^{\alpha}}{2} S_{\alpha}(S_{\alpha}+1)-\gamma B S_{\alpha} \\
\displaystyle
& = \frac{c^{\alpha}}{2} (S_{\alpha} - \frac{\gamma B}{c^{\alpha}} +\frac{1}{2} )^2 + const,
\end{array}
\end{equation}
for $\alpha= a,b$. Throughout the paper, we use $const$ to represent a constant whose actual value is not concerned and may not be the same each time it appears.

If $c^{\alpha} <0$, we always have $S_{\alpha}^m = N_{\alpha}$.
If $c^{\alpha} >0$, one finds the following three subcases. (i) If $c^{\alpha} \geq 2\gamma B$, then $S_{\alpha}^m =0$. (ii) If $\frac{2\gamma B}{2N_{\alpha}+1} \leq c^{\alpha} \leq 2\gamma B$, then $S_{\alpha}^m = n_{\alpha }\equiv Int(\frac{\gamma B}{c^{\alpha}}-\frac{1}{2})$, where $Int(x)$ denotes the integer closest to $x$ while in its legitimate range, e.g. $0 \leq n_{\alpha} \leq N_{\alpha}$.  (iii)  If $ c^{\alpha} \leq \frac{2\gamma B}{2N_{\alpha}+1} $, then $S_{\alpha}^m = N_{\alpha}.$ As $n_{\alpha}$ reduces to $0$ and $N_{\alpha}$ respectively at the two boundaries,  the ground state in $\frac{2\gamma B}{2N_{\alpha}+1} \leq c^{\alpha} \leq 2\gamma B$ is continuously connected with those in $c^{\alpha} \geq 2\gamma B$ and in $ c^{\alpha} \leq \frac{2\gamma B}{2N_{\alpha}+1} $.

\begin{figure*}
\setlength{\unitlength}{0.240900pt}
\ifx\plotpoint\undefined\newsavebox{\plotpoint}\fi
\begin{picture}(1500,900)(0,0)
\sbox{\plotpoint}{\rule[-0.200pt]{0.400pt}{0.400pt}}%
\put(330,664){\makebox(0,0)[l]{$|N_a,N_a
\rangle_a|0,0
\rangle_{b}$}}
\put(330,469){\makebox(0,0)[l]{$|N_a,N_a
\rangle_a|n_b,n_b
\rangle_{b}$}}
\put(330,274){\makebox(0,0)[l]{$|N_a,N_a
\rangle_a|N_b,N_b
\rangle_{b}$}}
\put(690,274){\makebox(0,0)[l]{$|n_a,n_a
\rangle_a|N_b,N_b
\rangle_{b}$}}
\put(690,664){\makebox(0,0)[l]{$|n_a,n_a
\rangle_a|0,0
\rangle_{b}$}}
\put(690,469){\makebox(0,0)[l]{$|n_a,n_a
\rangle_a|n_b,n_b
\rangle_{b}$}}
\put(1040,274){\makebox(0,0)[l]{$|0,0
\rangle_a|N_b,N_b
\rangle_{b}$}}
\put(1040,664){\makebox(0,0)[l]{$|0,0
\rangle_a|0,0
\rangle_{b}$}}
\put(1040,664){\makebox(0,0)[l]{$|0,0
\rangle_a|0,0
\rangle_{b}$}}
\put(1040,469){\makebox(0,0)[l]{$|0,0
\rangle_a|n_b,n_b
\rangle_{b}$}}
\put(330,373){\line(1,0){1000}}
\put(330,606){\line(1,0){1000}}
\put(670,160){\line(0,1){600}}
\put(1020,160){\line(0,1){600}}
\end{picture}
\caption{\label{b0} Ground states in $c^a-c^b$ parameter plane for $c^{ab} = 0$ and $B>0$. They are all direct products of the ground states of the two species, and  are thus written in the form of $|S_a^m,S_a^m\rangle_a|S_b^m,S_b^m\rangle_b$.  The two horizontal lines are  $c^a = \frac{2\gamma B}{2N_a+1}$ and   $c^a = 2\gamma B$, while the two vertical lines are   $c^b = \frac{2\gamma B}{2N_b+1}$ and   $c^b = 2\gamma B$.  $n_a \equiv Int (\frac{\gamma B}{c^a}-\frac{1}{2})$,  $n_b \equiv Int (\frac{\gamma B}{c^b}-\frac{1}{2})$. }
\end{figure*}

Thus one obtains all the ground states in the parameter subspace of  $c^{ab}=0$ and $B > 0$, which can be written in the form of  $|S_a^m,S_a^m\rangle_a|S_b^m,S_b^m\rangle_b$ and as depicted in FIG.~\ref{b0}. There are nine regimes, each is defined  by the range of $c^a$ and $c^b$ specified above. In each regime, the ground state is a direct product of the ground states of the two species given above accordingly. Each ground state is continuously connected with those in the neighboring regimes.
Therefore, on $c^a-c^b$ plane, ground states in all regimes belong to a same quantum phase.

As $B \rightarrow 0$, however, the five crossover regimes tend to vanish, and the four ground states in the remaining four corner regimes become discontinuous, as already known~\cite{shi5}.
Therefore, the quantum phase transitions among the ground states in the four quadrants of $c^a-c^b$ plane for $c^{ab} = 0$ in absence of a magnetic field can be circumvented by turning on and then off a magnetic field. Hence a magnetic field has an interesting effect even in the regime without interspecies spin exchange.

One can imagine the three-dimensional parameter subspace of $c^{ab}=0$, with $c^a$, $c^b$ and $B\geq 0$ as the three coordinates. The boundaries $c^{a}=\frac{2\gamma B}{2N_{a}+1}$, $c^a= 2\gamma B$, $c^{b}=\frac{2\gamma B}{2N_{b}+1}$ and $c^b= 2\gamma B$ are all planes starting from the origin and extending to positive infinities.

\section{$c^{ab} < 0$ \label{ferro}}

For $c^{ab} \neq 0$, we have worked out the complicated problem of minimizing $E$ with four variables $S_a$, $S_b$, $S$ and $B$ in most parameter regimes. But in some regimes,  the calculations are too difficult or complicated for us to obtain the results. The calculation details are given in Appendix~A. The ground states we obtained are listed in Table~\ref{table}.

\begin{table*}
\begin{tabular}{|l|l|l|l|l|}
\hline No. & \multicolumn{3}{|c|}{Parameter regimes} & {Ground states} \\
 \hline
 \multirow{2}{*}{1,A2a} &  & \multicolumn{2}{|c|}{ $c^{a} \leq 0$,}
 & $|N_a,N_a\rangle_a|N_b,N_b\rangle_b$,\\
 &&\multicolumn{2}{|c|}{ $c^{b}\leq \frac{2\gamma B -2N_ac^{ab}}{2N_b+1}$}& disentangled \\ \cline{1-1} \cline{3-5}
 \multirow{2}{*}{A2b} & & \multicolumn{2}{|c|}{ $c^{a} \leq 0$,} &
 $|N_a,N_a\rangle_a|n_1',n_1'\rangle_b$ \\
 &&\multicolumn{2}{|c|}{ $\frac{2\gamma B -2N_ac^{ab}}{2N_b+1}\leq c_b \leq 2\gamma B  -2N_ac^{ab}$} & disentangled,
 $n_1' \equiv Int[\frac{\gamma B+N_a|c^{ab}|}{c^{b}}-\frac{1}{2}]$ \\\cline{1-1} \cline{3-5}
 \multirow{2}{*}{A2c}&  $c^{ab} < 0$ &\multicolumn{2}{|c|}{  $c^{a} \leq 0$,}
 & $|N_a, S_z\rangle_a|0,0\rangle_b$,\\
 &&\multicolumn{2}{|c|}{ $c^{b}\geq 2\gamma B -2N_ac^{ab} $}& disentangled  \\ \cline{1-1} \cline{3-5}
  \multirow{2}{*}{1,A3a}&  & \multicolumn{2}{|c|}{ $c^{b} \leq 0$,}
 &  $|N_a,N_a\rangle_a|N_b,N_b\rangle_b$,\\
&&\multicolumn{2}{|c|}{ $c^{a}\leq \frac{2\gamma B -2N_bc^{ab}}{2N_a+1}$}& disentangled  \\ \cline{1-1} \cline{3-5}
 \multirow{2}{*}{A3b} &  &\multicolumn{2}{|c|}{  $c^{b} \leq 0$,} &
 $|n_2',n_2'\rangle_a|N_b, N_b\rangle_b$\\
 & &\multicolumn{2}{|c|}{ $\frac{2\gamma B -2N_bc^{ab}}{2N_a+1} \leq c_a \leq 2\gamma B -2N_bc^{ab}$ }& disentangled,
 $n_2' \equiv Int[\frac{\gamma B+N_b|c^{ab}|}{c^{a}}-\frac{1}{2}]$ \\\cline{1-1} \cline{3-5}
  \multirow{2}{*}{A3c} &  & \multicolumn{2}{|c|}{ $c^{b} \leq 0$,}
 & $|0,0\rangle_a|N_b,N_b\rangle_b$,\\
 &   &\multicolumn{2}{|c|}{ $c^{a}\geq 2\gamma B -2N_bc^{ab} $}& disentangled \\ \cline{1-1}\cline{3-5}
 \hline
  \multirow{2}{*}{IV}  & & \multicolumn{2}{|c|}{ $c^{a} < c^{ab}$,  $c^{b} < c^{ab}$}& $|N_a,N_b,n,n\rangle$,  $n \equiv Int(\frac{\gamma B
}{c^{ab} }- \frac{1}{2})$\\
 && \multicolumn{2}{|c|}{(boundaries were discussed in Ref.~\cite{shi4})} &  entangled, \\
 \cline{1-1} \cline{3-5}
  \multirow{2}{*}{B1} & $0< c^{ab} \leq 2\gamma B$  & &
  \multirow{2}{*}{$c^{b}\geq 2\gamma B$} & $|0,0\rangle_a|0,0\rangle_b$,\\
  &&&&disentangled \\ \cline{1-1} \cline{4-5}
  \multirow{2}{*}{B2a}
& & $c^{a}\geq c^{b}>c^{ab}$&
   $c^{b}<2\gamma B\leq {\frac{c^{a}-c^{ab}}{c^{b}-c^{ab}}}c^{b}$
 & $|0,0\rangle_a  |n_b,n_b\rangle_b$, $n_b \equiv Int[\frac{\gamma
B}{c^{b}}-\frac{1}{2}]$ \\ &&$N_a=N_b=N$&&disentangled \\ \cline{1-1} \cline{4-5}  \multirow{2}{*}{B2b}
 &  &  & $2\gamma B>\frac{c^{a}-c^{ab}}{c^{b}-c^{ab}}c^{b}$
 & $|n_{a3},n_{a3}\rangle_a  |n_{b1},n_{b1}\rangle_b$, $n_{a3} \equiv Int[\frac{2\gamma
B|c^{b}-c^{ab}|-c^{b}(c^{a}-c^{ab})}{2[c^{a}c^{b}-(c^{ab})^2]}]$\\
 & & & & disentangled, $n_{b1}\equiv Int[\frac{\gamma
B-c^{ab}n_{a3}}{c^{b}}-\frac{1}{2}]$ \\ \cline{1-1}
 \cline{3-5} \multirow{2}{*}{C1}
 & &  $ c^{a} >  c^{ab} > c^{b}$ & \multirow{2}{*}{ $2\gamma B\leq{\frac{c^{a}-c^{ab}}{c^{ab}-c^{b}}c^{b}}$}
 & $|0,n_b,n_b,n_b\rangle=|0,0\rangle_a   |n_b,n_b\rangle_b$
 \\
 &&$c^{a}c^{b} > {(c^{ab})^2}$ & &disentangled
 \\  \cline{1-1} \cline{4-5}\multirow{2}{*}{C2}
 &  &$N_a=N_b=N >N^*$
 & \multirow{2}{*}{ $2\gamma B>\frac{c^{a}-c^{ab}}{c^{ab}-c^{b}}c^{b}$}
 & $|n_{a3},n_{b2},n_{b2}-n_{a3},n_{b2}-n_{a3}\rangle$\\
 &&
 && entangled,
$n_{b2} \equiv Int[\frac{\gamma
B+c^{ab}n_{a3}}{c^{b}}-\frac{1}{2}]$ \\\cline{1-1} \cline{3-5}
  \multirow{2}{*}{D1} &   & &
  \multirow{2}{*}{$c^{a}\geq 2\gamma B$} & $|0,0\rangle_a|0,0\rangle_b$,\\
  &&&&disentangled \\ \cline{1-1} \cline{4-5}
  \multirow{2}{*}{D2a}
& & $c^{b}\geq c^{a}>c^{ab}$&
   $c^{a}<2\gamma B\leq {\frac{c^{b}-c^{ab}}{c^{a}-c^{ab}}}c^{a}$
 & $ |n_a,n_a\rangle_a |0,0\rangle_b $, $n_a\equiv Int[\frac{\gamma
B}{c^{a}}-\frac{1}{2}]$ \\ &&$N_a=N_b=N$&&disentangled \\ \cline{1-1} \cline{4-5}  \multirow{2}{*}{D2b}
 &  &  & $2\gamma B>\frac{c^{b}-c^{ab}}{c^{a}-c^{ab}}c^{a}$
 & $ |n_{a1},n_{a1}\rangle_a|n_{b3},n_{b3}\rangle_b$, $n_{b3} \equiv Int[\frac{2\gamma
B|c^{a}-c^{ab}|-c^{a}(c^{b}-c^{ab})}{2[c^{a}c^{b}-(c^{ab})^2]}]$\\
 & & & & disentangled,  $n_{a1}\equiv Int[\frac{\gamma
B-c^{ab}n_{b3}}{c^{a}}-\frac{1}{2}]$ \\ \cline{1-1}
 \cline{3-5} \multirow{2}{*}{E1}
 & &  $ c^{b} >  c^{ab} > c^{a}$ & \multirow{2}{*}{ $2\gamma B\leq{\frac{c^{b}-c^{ab}}{c^{ab}-c^{a}}c^{a}}$}
 & $|n_a,0,n_a,n_a\rangle=|n_a,n_a\rangle_a|0,0\rangle_b  $
 \\
 &&$c^{a}c^{b} > {(c^{ab})^2}$ & &disentangled
 \\  \cline{1-1} \cline{4-5}\multirow{2}{*}{E2}
 &  &$N_a=N_b=N >N^*$
 & \multirow{2}{*}{ $2\gamma B>\frac{c^{b}-c^{ab}}{c^{ab}-c^{a}}c^{a}$}
 & $|n_{a2},n_{b3},n_{a2}-n_{b3},n_{a2}-n_{b3}\rangle$\\
 &&&& entangled,
$n_{a2} \equiv Int[\frac{\gamma
B+c^{ab}n_{b3}}{c^{a}}-\frac{1}{2}]$ \\
 \hline
 \multirow{2}{*}{II}
     & \multirow{4}{*}{$c^{ab} \geq  2\gamma B >0 $}  &  \multicolumn{2}{|c|}{$c^{a} > c^{ab}$, $c^{b} > c^{ab}$}  & $|0,0\rangle_a|0,0\rangle_b$,   \\
 & &\multicolumn{2}{|c|}{(boundaries were discussed in Ref.~\cite{shi4})} &  disentangled \\ \cline{1-1}
   \cline{3-5}
\multirow{2}{*}{III} &  & \multicolumn{2}{|c|}{$c^{a} < c^{ab}$, $c^{b} < c^{ab}$,  $N_a=N_b=N$ } & $|N,N,0,0\rangle $ \\
 &  & \multicolumn{2}{|c|}{(boundaries were discussed in Ref.~\cite{shi4})}& entangled \\
 \hline
 \end{tabular}
\caption{\label{table} Ground states of a mixture of two spin-1 atomic gases in different parameter regimes for $c^{ab} \neq 0$ and $B > 0$. For $c^{ab} <0$, the ground states are always disentangled, and are written in the form of $|S_a^m,S_a^m\rangle_a|S_b^m,S_b^m\rangle_b$. For $c^{ab} > 0$, some ground states are disentangled and are also written in this form. Some ground states are written in the form of $|S_a^m,S_a^m,S^m,S^m\rangle$, which may be entangled or disentangled, depending on the parameter values. The ordering numbers are just the corresponding section numbers in the Appendices where the calculation are done,  or  as numbered  in Ref.~\cite{shi4a} for those calculated there (IV, II and III).
 $N^*$ is defined in Eq.~(\ref{nn}). }
\end{table*}

\begin{figure*}
\setlength{\unitlength}{0.240900pt}
\ifx\plotpoint\undefined\newsavebox{\plotpoint}\fi
\begin{picture}(1500,900)(0,0)
\sbox{\plotpoint}{\rule[-0.200pt]{0.400pt}{0.400pt}}%
\put(1242,340){\makebox(0,0)[l]{$c^a$}}
\put(600,859){\makebox(0,0)[l]{$c^b$}}
\put(330,664){\makebox(0,0)[l]{$|N_a,N_a
\rangle_a|0,0
\rangle_{b}$}}
\put(330,469){\makebox(0,0)[l]{$|N_a,N_a
\rangle_a|n_1',n_1'
\rangle_{b}$}}
\put(330,274){\makebox(0,0)[l]{$|N_a,N_a
\rangle_a|N_b,N_b
\rangle_{b}$}}
\put(668,274){\makebox(0,0)[l]{$|n_2',n_2'
\rangle_a|N_b,N_b
\rangle_{b}$}}
\put(993,274){\makebox(0,0)[l]{$|0,0
\rangle_a|N_b,N_b
\rangle_{b}$}}
\put(642,352){\vector(1,0){580}}
\put(642,352){\vector(0,1){486}}
\put(330,373){\line(1,0){312}}
\put(330,606){\line(1,0){312}}
\put(662,40){\line(0,1){312}}
\put(970,40){\line(0,1){312}}
\put(642,373){\line(0,1){486}}
\put(662,352){\line(1,0){560}}
\end{picture}
\caption{Ground states in $c^a~-~c^b$ parameter plane for $c^{ab} <0$ and $B >0$. They are all disentangled and are thus written in the form of $|S_a^m,S_a^m\rangle_a|S_b^m,S_b^m\rangle_b$.  The states in the second, third and fourth quadrants are all continuously connected on boundaries. $n_1' \equiv Int[\frac{\gamma B+N_a|c^{ab}}{c^{b}}-\frac{1}{2}]$, $n_2' \equiv Int[\frac{\gamma B+N_b|c^{ab}|}{c^{a}}-\frac{1}{2}]$. The four boundaries are $c_a = \frac{2\gamma B -2N_bc^{ab}}{2N_a+1}$, $c_a = 2\gamma B -2N_bc^{ab}$, $c_b =\frac{2\gamma B -2N_ac^{ab}}{2N_b+1}$ and  $c_b = 2\gamma B  -2N_ac^{ab}$.  The states in the first quadrant have not yet been determined. \label{b1}}
\end{figure*}
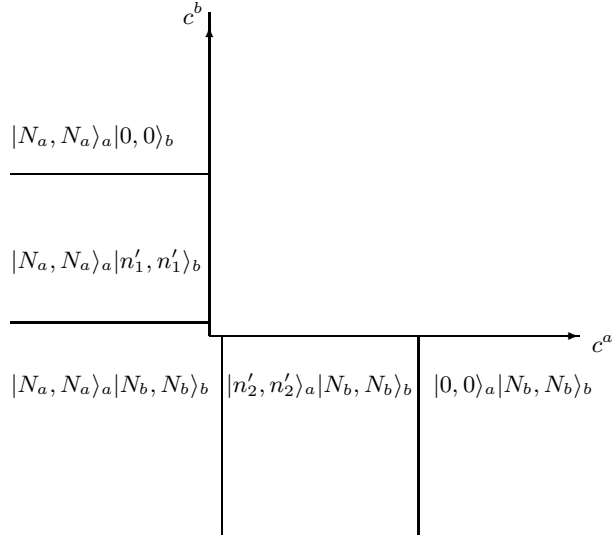

For $c^{ab} < 0$ while $B>0$, we have found the ground states in the second, third and fourth quadrants of $c^a-c^b$ plane, as depicted in FIG.~\ref{b1}. In comparison with the case of $B=0$~\cite{shi5}, a magnetic field $B>0$ both shifts the positions of the boundaries and modifies the ground states in the crossover regimes.

In the three outmost regimes, only the boundaries are shifted, while the ground states remain the same as those of $B=0$.  The ground state is $|N_a,N_a\rangle_a|N_b,N_b\rangle_b$ in the regime $c^{a} \leq 0$ while $c^{b}\leq \frac{2\gamma B -2N_ac^{ab}}{2N_b+1}$ and  $c^{b} \leq 0$ while $c^{a}\leq \frac{2\gamma B -2N_bc^{ab}}{2N_a+1}$. In the regimes  $c^{a} \leq 0$ while $c^{b}\geq 2\gamma B -2N_ac^{ab} $ and  $c^{b} \leq 0$ while $c^{a}\geq 2\gamma B -2N_bc^{ab} $, the ground states are $|N_a, N_a\rangle_a|0,0\rangle_b$ and $|0,0\rangle_a|N_b,N_b\rangle_b$ respectively.

In the crossover regimes, the ground states are also modified. For
$c^{a} \leq 0$ while $\frac{2\gamma B -2N_ac^{ab}}{2N_b+1}\leq c_b \leq 2\gamma B  -2N_ac^{ab}$, the ground state is  $|N_a,N_a\rangle_a|n_1',n_1'\rangle_b$, where  $n_1'=Int[\frac{\gamma B+N_a|c^{ab}}{c^{b}}-\frac{1}{2}]$. Likewise, for  $c^{b} \leq 0$ while $\frac{2\gamma B -2N_bc^{ab}}{2N_a+1} \leq c_a \leq 2\gamma B -2N_bc^{ab}$, the ground state is  $|n_2',n_2'\rangle_a|N_b, N_b\rangle_b$, where $n_2'=Int[\frac{\gamma B+N_b|c^{ab}|}{c^{a}}-\frac{1}{2}]$.

We see continuous connections in both $c^{ab}$ and $B$ dimensions. As $B\rightarrow 0$, all the ground states in these three quadrants of $c^a-c^b$ plane for $c^{ab} <0$ reduce to the corresponding ones in absence of a magnetic field. On the other hand, as $c^{ab} \rightarrow 0$, the ground states in the these three quadrants reduce to those for $c^{ab}=0$, given in last section.

Note that in all subregimes of $c^{ab} \leq 0$, the ground states are always disentangled, as $S^m=S_a^m+S_b^m$.

\section{$0<c^{ab}< 2\gamma B$ \label{anti1}}

Now we turn to antiferromagnetic interspecies spin coupling.
For $0<c^{ab} \leq 2\gamma B$, it has been known previously that if $c^{a} < c^{ab}$ and  $c^{b} < c^{ab}$, the ground state is $|N_a,N_b,n,n\rangle$, where  $n \equiv Int(\frac{\gamma B
}{c^{ab} }- \frac{1}{2})$ satisfies $|N_a-N_b| \leq n \leq N_a+N_b$~\cite{shi5}. This state is entangled unless $n=N_a+N_b$.  The ground states on the  boundaries $c^a=c^{ab}$ and $c^b=c^{ab}$ have also been discussed in details. Especially,  it has been known that if $c^{a}=c^{b}=c^{ab}$, then there are many degenerate ground states in the form of $|S_a,S_b,n,n\rangle$, as far as $S_a$, $S_b$ and $n$ satisfy the constraint $|S_a-S_b| \leq n \leq S_a+S_b$.

\begin{figure*}
\setlength{\unitlength}{0.240900pt}
\ifx\plotpoint\undefined\newsavebox{\plotpoint}\fi
\sbox{\plotpoint}{\rule[-0.200pt]{0.400pt}{1.400pt}}%
\begin{picture}(1500,900)(0,0)
\sbox{\plotpoint}{\rule[-0.200pt]{0.400pt}{1.400pt}}%
\put(330,229){\line(1,0){189}}
\put(519,40){\line(0,1){189}}
\put(519,229){\line(1,1){252}}
\put(771,481){\line(1,0){378}}
\put(771,481){\line(0,1){378}}
\put(320,150){\makebox(0,0)[l]{$|N,N,n,n\rangle$}}
\put(800,648){\makebox(0,0)[l]{$|0,0\rangle_a|0,0\rangle_b$}}
\put(590,380){\makebox(0,0)[l]{\small D2b}}
\put(620,330){\makebox(0,0)[l]{\small B2b}}
\put(408,470){\makebox(0,0)[l]{\small E2}}
\put(650,170){\makebox(0,0)[l]{\small C2}}
\put(800,348){\makebox(0,0)[l]{$|0,0\rangle_a|n_b,n_b\rangle_b$}}
\put(500,648){\makebox(0,0)[l]{$|n_a,n_a\rangle_a|0,0\rangle_b$}}
\multiput(414.59,813.75)(0.485,-14.375){11}{\rule{0.117pt}{10.900pt}}
\multiput(413.17,836.38)(7.000,-166.377){2}{\rule{0.400pt}{5.450pt}}
\multiput(421.59,642.81)(0.488,-8.484){13}{\rule{0.117pt}{6.550pt}}
\multiput(420.17,656.41)(8.000,-115.405){2}{\rule{0.400pt}{3.275pt}}
\multiput(429.59,525.46)(0.489,-4.747){15}{\rule{0.118pt}{3.744pt}}
\multiput(428.17,533.23)(9.000,-74.228){2}{\rule{0.400pt}{1.872pt}}
\multiput(438.59,447.17)(0.488,-3.597){13}{\rule{0.117pt}{2.850pt}}
\multiput(437.17,453.08)(8.000,-49.085){2}{\rule{0.400pt}{1.425pt}}
\multiput(446.59,395.08)(0.488,-2.673){13}{\rule{0.117pt}{2.150pt}}
\multiput(445.17,399.54)(8.000,-36.538){2}{\rule{0.400pt}{1.075pt}}
\multiput(454.59,356.15)(0.488,-2.013){13}{\rule{0.117pt}{1.650pt}}
\multiput(453.17,359.58)(8.000,-27.575){2}{\rule{0.400pt}{0.825pt}}
\multiput(462.59,326.97)(0.489,-1.427){15}{\rule{0.118pt}{1.211pt}}
\multiput(461.17,329.49)(9.000,-22.486){2}{\rule{0.400pt}{0.606pt}}
\multiput(471.59,302.64)(0.488,-1.220){13}{\rule{0.117pt}{1.050pt}}
\multiput(470.17,304.82)(8.000,-16.821){2}{\rule{0.400pt}{0.525pt}}
\multiput(479.59,284.26)(0.488,-1.022){13}{\rule{0.117pt}{0.900pt}}
\multiput(478.17,286.13)(8.000,-14.132){2}{\rule{0.400pt}{0.450pt}}
\multiput(487.59,268.68)(0.488,-0.890){13}{\rule{0.117pt}{0.800pt}}
\multiput(486.17,270.34)(8.000,-12.340){2}{\rule{0.400pt}{0.400pt}}
\multiput(495.59,255.37)(0.489,-0.669){15}{\rule{0.118pt}{0.633pt}}
\multiput(494.17,256.69)(9.000,-10.685){2}{\rule{0.400pt}{0.317pt}}
\multiput(504.59,243.51)(0.488,-0.626){13}{\rule{0.117pt}{0.600pt}}
\multiput(503.17,244.75)(8.000,-8.755){2}{\rule{0.400pt}{0.300pt}}
\multiput(512.00,234.93)(0.494,-0.488){13}{\rule{0.500pt}{0.117pt}}
\multiput(512.00,235.17)(6.962,-8.000){2}{\rule{0.250pt}{0.400pt}}
\multiput(520.00,226.93)(0.560,-0.488){13}{\rule{0.550pt}{0.117pt}}
\multiput(520.00,227.17)(7.858,-8.000){2}{\rule{0.275pt}{0.400pt}}
\multiput(529.00,218.93)(0.569,-0.485){11}{\rule{0.557pt}{0.117pt}}
\multiput(529.00,219.17)(6.844,-7.000){2}{\rule{0.279pt}{0.400pt}}
\multiput(537.00,211.93)(0.671,-0.482){9}{\rule{0.633pt}{0.116pt}}
\multiput(537.00,212.17)(6.685,-6.000){2}{\rule{0.317pt}{0.400pt}}
\multiput(545.00,205.93)(0.821,-0.477){7}{\rule{0.740pt}{0.115pt}}
\multiput(545.00,206.17)(6.464,-5.000){2}{\rule{0.370pt}{0.400pt}}
\multiput(553.00,200.93)(0.933,-0.477){7}{\rule{0.820pt}{0.115pt}}
\multiput(553.00,201.17)(7.298,-5.000){2}{\rule{0.410pt}{0.400pt}}
\multiput(562.00,195.94)(1.066,-0.468){5}{\rule{0.900pt}{0.113pt}}
\multiput(562.00,196.17)(6.132,-4.000){2}{\rule{0.450pt}{0.400pt}}
\multiput(570.00,191.94)(1.066,-0.468){5}{\rule{0.900pt}{0.113pt}}
\multiput(570.00,192.17)(6.132,-4.000){2}{\rule{0.450pt}{0.400pt}}
\multiput(578.00,187.94)(1.066,-0.468){5}{\rule{0.900pt}{0.113pt}}
\multiput(578.00,188.17)(6.132,-4.000){2}{\rule{0.450pt}{0.400pt}}
\multiput(586.00,183.95)(1.802,-0.447){3}{\rule{1.300pt}{0.108pt}}
\multiput(586.00,184.17)(6.302,-3.000){2}{\rule{0.650pt}{0.400pt}}
\multiput(595.00,180.95)(1.579,-0.447){3}{\rule{1.167pt}{0.108pt}}
\multiput(595.00,181.17)(5.579,-3.000){2}{\rule{0.583pt}{0.400pt}}
\multiput(603.00,177.95)(1.579,-0.447){3}{\rule{1.167pt}{0.108pt}}
\multiput(603.00,178.17)(5.579,-3.000){2}{\rule{0.583pt}{0.400pt}}
\multiput(611.00,174.95)(1.802,-0.447){3}{\rule{1.300pt}{0.108pt}}
\multiput(611.00,175.17)(6.302,-3.000){2}{\rule{0.650pt}{0.400pt}}
\put(620,171.17){\rule{1.700pt}{0.400pt}}
\multiput(620.00,172.17)(4.472,-2.000){2}{\rule{0.850pt}{0.400pt}}
\multiput(628.00,169.95)(1.579,-0.447){3}{\rule{1.167pt}{0.108pt}}
\multiput(628.00,170.17)(5.579,-3.000){2}{\rule{0.583pt}{0.400pt}}
\put(636,166.17){\rule{1.700pt}{0.400pt}}
\multiput(636.00,167.17)(4.472,-2.000){2}{\rule{0.850pt}{0.400pt}}
\put(644,164.17){\rule{1.900pt}{0.400pt}}
\multiput(644.00,165.17)(5.056,-2.000){2}{\rule{0.950pt}{0.400pt}}
\put(653,162.17){\rule{1.700pt}{0.400pt}}
\multiput(653.00,163.17)(4.472,-2.000){2}{\rule{0.850pt}{0.400pt}}
\put(661,160.17){\rule{1.700pt}{0.400pt}}
\multiput(661.00,161.17)(4.472,-2.000){2}{\rule{0.850pt}{0.400pt}}
\put(669,158.67){\rule{1.927pt}{0.400pt}}
\multiput(669.00,159.17)(4.000,-1.000){2}{\rule{0.964pt}{0.400pt}}
\put(677,157.17){\rule{1.900pt}{0.400pt}}
\multiput(677.00,158.17)(5.056,-2.000){2}{\rule{0.950pt}{0.400pt}}
\put(686,155.67){\rule{1.927pt}{0.400pt}}
\multiput(686.00,156.17)(4.000,-1.000){2}{\rule{0.964pt}{0.400pt}}
\put(694,154.17){\rule{1.700pt}{0.400pt}}
\multiput(694.00,155.17)(4.472,-2.000){2}{\rule{0.850pt}{0.400pt}}
\put(702,152.67){\rule{2.168pt}{0.400pt}}
\multiput(702.00,153.17)(4.500,-1.000){2}{\rule{1.084pt}{0.400pt}}
\put(711,151.67){\rule{1.927pt}{0.400pt}}
\multiput(711.00,152.17)(4.000,-1.000){2}{\rule{0.964pt}{0.400pt}}
\put(719,150.67){\rule{1.927pt}{0.400pt}}
\multiput(719.00,151.17)(4.000,-1.000){2}{\rule{0.964pt}{0.400pt}}
\put(727,149.17){\rule{1.700pt}{0.400pt}}
\multiput(727.00,150.17)(4.472,-2.000){2}{\rule{0.850pt}{0.400pt}}
\put(735,147.67){\rule{2.168pt}{0.400pt}}
\multiput(735.00,148.17)(4.500,-1.000){2}{\rule{1.084pt}{0.400pt}}
\put(744,146.67){\rule{1.927pt}{0.400pt}}
\multiput(744.00,147.17)(4.000,-1.000){2}{\rule{0.964pt}{0.400pt}}
\put(752,145.67){\rule{1.927pt}{0.400pt}}
\multiput(752.00,146.17)(4.000,-1.000){2}{\rule{0.964pt}{0.400pt}}
\put(760,144.67){\rule{1.927pt}{0.400pt}}
\multiput(760.00,145.17)(4.000,-1.000){2}{\rule{0.964pt}{0.400pt}}
\put(768,143.67){\rule{2.168pt}{0.400pt}}
\multiput(768.00,144.17)(4.500,-1.000){2}{\rule{1.084pt}{0.400pt}}
\put(777,142.67){\rule{1.927pt}{0.400pt}}
\multiput(777.00,143.17)(4.000,-1.000){2}{\rule{0.964pt}{0.400pt}}
\put(793,141.67){\rule{2.168pt}{0.400pt}}
\multiput(793.00,142.17)(4.500,-1.000){2}{\rule{1.084pt}{0.400pt}}
\put(802,140.67){\rule{1.927pt}{0.400pt}}
\multiput(802.00,141.17)(4.000,-1.000){2}{\rule{0.964pt}{0.400pt}}
\put(810,139.67){\rule{1.927pt}{0.400pt}}
\multiput(810.00,140.17)(4.000,-1.000){2}{\rule{0.964pt}{0.400pt}}
\put(785.0,143.0){\rule[-0.200pt]{1.927pt}{0.400pt}}
\put(826,138.67){\rule{2.168pt}{0.400pt}}
\multiput(826.00,139.17)(4.500,-1.000){2}{\rule{1.084pt}{0.400pt}}
\put(835,137.67){\rule{1.927pt}{0.400pt}}
\multiput(835.00,138.17)(4.000,-1.000){2}{\rule{0.964pt}{0.400pt}}
\put(818.0,140.0){\rule[-0.200pt]{1.927pt}{0.400pt}}
\put(851,136.67){\rule{1.927pt}{0.400pt}}
\multiput(851.00,137.17)(4.000,-1.000){2}{\rule{0.964pt}{0.400pt}}
\put(859,135.67){\rule{2.168pt}{0.400pt}}
\multiput(859.00,136.17)(4.500,-1.000){2}{\rule{1.084pt}{0.400pt}}
\put(843.0,138.0){\rule[-0.200pt]{1.927pt}{0.400pt}}
\put(876,134.67){\rule{1.927pt}{0.400pt}}
\multiput(876.00,135.17)(4.000,-1.000){2}{\rule{0.964pt}{0.400pt}}
\put(868.0,136.0){\rule[-0.200pt]{1.927pt}{0.400pt}}
\put(893,133.67){\rule{1.927pt}{0.400pt}}
\multiput(893.00,134.17)(4.000,-1.000){2}{\rule{0.964pt}{0.400pt}}
\put(884.0,135.0){\rule[-0.200pt]{2.168pt}{0.400pt}}
\put(909,132.67){\rule{1.927pt}{0.400pt}}
\multiput(909.00,133.17)(4.000,-1.000){2}{\rule{0.964pt}{0.400pt}}
\put(901.0,134.0){\rule[-0.200pt]{1.927pt}{0.400pt}}
\put(926,131.67){\rule{1.927pt}{0.400pt}}
\multiput(926.00,132.17)(4.000,-1.000){2}{\rule{0.964pt}{0.400pt}}
\put(917.0,133.0){\rule[-0.200pt]{2.168pt}{0.400pt}}
\put(942,130.67){\rule{1.927pt}{0.400pt}}
\multiput(942.00,131.17)(4.000,-1.000){2}{\rule{0.964pt}{0.400pt}}
\put(934.0,132.0){\rule[-0.200pt]{1.927pt}{0.400pt}}
\put(967,129.67){\rule{1.927pt}{0.400pt}}
\multiput(967.00,130.17)(4.000,-1.000){2}{\rule{0.964pt}{0.400pt}}
\put(950.0,131.0){\rule[-0.200pt]{4.095pt}{0.400pt}}
\put(992,128.67){\rule{1.927pt}{0.400pt}}
\multiput(992.00,129.17)(4.000,-1.000){2}{\rule{0.964pt}{0.400pt}}
\put(975.0,130.0){\rule[-0.200pt]{4.095pt}{0.400pt}}
\put(1008,127.67){\rule{2.168pt}{0.400pt}}
\multiput(1008.00,128.17)(4.500,-1.000){2}{\rule{1.084pt}{0.400pt}}
\put(1000.0,129.0){\rule[-0.200pt]{1.927pt}{0.400pt}}
\put(1033,126.67){\rule{1.927pt}{0.400pt}}
\multiput(1033.00,127.17)(4.000,-1.000){2}{\rule{0.964pt}{0.400pt}}
\put(1017.0,128.0){\rule[-0.200pt]{3.854pt}{0.400pt}}
\put(1066,125.67){\rule{2.168pt}{0.400pt}}
\multiput(1066.00,126.17)(4.500,-1.000){2}{\rule{1.084pt}{0.400pt}}
\put(1041.0,127.0){\rule[-0.200pt]{6.022pt}{0.400pt}}
\put(1091,124.67){\rule{1.927pt}{0.400pt}}
\multiput(1091.00,125.17)(4.000,-1.000){2}{\rule{0.964pt}{0.400pt}}
\put(1075.0,126.0){\rule[-0.200pt]{3.854pt}{0.400pt}}
\put(1124,123.67){\rule{1.927pt}{0.400pt}}
\multiput(1124.00,124.17)(4.000,-1.000){2}{\rule{0.964pt}{0.400pt}}
\put(1099.0,125.0){\rule[-0.200pt]{6.022pt}{0.400pt}}
\put(1132.0,124.0){\rule[-0.200pt]{4.095pt}{0.400pt}}
\put(531.84,233.07){\usebox{\plotpoint}}
\put(550.99,241.00){\usebox{\plotpoint}}
\put(570.08,249.03){\usebox{\plotpoint}}
\put(588.93,257.63){\usebox{\plotpoint}}
\put(607.12,267.58){\usebox{\plotpoint}}
\put(624.99,278.12){\usebox{\plotpoint}}
\put(641.77,290.33){\usebox{\plotpoint}}
\put(658.38,302.71){\usebox{\plotpoint}}
\put(673.70,316.70){\usebox{\plotpoint}}
\put(688.53,331.16){\usebox{\plotpoint}}
\put(701.49,347.37){\usebox{\plotpoint}}
\put(714.60,363.40){\usebox{\plotpoint}}
\put(725.72,380.92){\usebox{\plotpoint}}
\put(735.64,399.15){\usebox{\plotpoint}}
\put(745.58,417.36){\usebox{\plotpoint}}
\put(754.11,436.27){\usebox{\plotpoint}}
\put(761.61,455.62){\usebox{\plotpoint}}
\sbox{\plotpoint}{\rule[-0.400pt]{0.800pt}{0.800pt}}%
\multiput(520.00,227.34)(4.500,-3.000){2}{\rule{1.084pt}{0.800pt}}
\put(529,222.84){\rule{1.927pt}{0.800pt}}
\multiput(529.00,224.34)(4.000,-3.000){2}{\rule{0.964pt}{0.800pt}}
\put(537,220.34){\rule{1.927pt}{0.800pt}}
\multiput(537.00,221.34)(4.000,-2.000){2}{\rule{0.964pt}{0.800pt}}
\put(545,217.84){\rule{1.927pt}{0.800pt}}
\multiput(545.00,219.34)(4.000,-3.000){2}{\rule{0.964pt}{0.800pt}}
\put(553,215.34){\rule{2.168pt}{0.800pt}}
\multiput(553.00,216.34)(4.500,-2.000){2}{\rule{1.084pt}{0.800pt}}
\put(562,213.34){\rule{1.927pt}{0.800pt}}
\multiput(562.00,214.34)(4.000,-2.000){2}{\rule{0.964pt}{0.800pt}}
\put(570,211.34){\rule{1.927pt}{0.800pt}}
\multiput(570.00,212.34)(4.000,-2.000){2}{\rule{0.964pt}{0.800pt}}
\put(578,209.34){\rule{1.927pt}{0.800pt}}
\multiput(578.00,210.34)(4.000,-2.000){2}{\rule{0.964pt}{0.800pt}}
\put(586,207.34){\rule{2.168pt}{0.800pt}}
\multiput(586.00,208.34)(4.500,-2.000){2}{\rule{1.084pt}{0.800pt}}
\put(595,205.34){\rule{1.927pt}{0.800pt}}
\multiput(595.00,206.34)(4.000,-2.000){2}{\rule{0.964pt}{0.800pt}}
\put(603,203.34){\rule{1.927pt}{0.800pt}}
\multiput(603.00,204.34)(4.000,-2.000){2}{\rule{0.964pt}{0.800pt}}
\put(611,201.84){\rule{2.168pt}{0.800pt}}
\multiput(611.00,202.34)(4.500,-1.000){2}{\rule{1.084pt}{0.800pt}}
\put(620,200.34){\rule{1.927pt}{0.800pt}}
\multiput(620.00,201.34)(4.000,-2.000){2}{\rule{0.964pt}{0.800pt}}
\put(628,198.34){\rule{1.927pt}{0.800pt}}
\multiput(628.00,199.34)(4.000,-2.000){2}{\rule{0.964pt}{0.800pt}}
\put(636,196.84){\rule{1.927pt}{0.800pt}}
\multiput(636.00,197.34)(4.000,-1.000){2}{\rule{0.964pt}{0.800pt}}
\put(644,195.34){\rule{2.168pt}{0.800pt}}
\multiput(644.00,196.34)(4.500,-2.000){2}{\rule{1.084pt}{0.800pt}}
\put(653,193.84){\rule{1.927pt}{0.800pt}}
\multiput(653.00,194.34)(4.000,-1.000){2}{\rule{0.964pt}{0.800pt}}
\put(661,192.34){\rule{1.927pt}{0.800pt}}
\multiput(661.00,193.34)(4.000,-2.000){2}{\rule{0.964pt}{0.800pt}}
\put(669,190.84){\rule{1.927pt}{0.800pt}}
\multiput(669.00,191.34)(4.000,-1.000){2}{\rule{0.964pt}{0.800pt}}
\put(677,189.34){\rule{2.168pt}{0.800pt}}
\multiput(677.00,190.34)(4.500,-2.000){2}{\rule{1.084pt}{0.800pt}}
\put(686,187.84){\rule{1.927pt}{0.800pt}}
\multiput(686.00,188.34)(4.000,-1.000){2}{\rule{0.964pt}{0.800pt}}
\put(694,186.84){\rule{1.927pt}{0.800pt}}
\multiput(694.00,187.34)(4.000,-1.000){2}{\rule{0.964pt}{0.800pt}}
\put(702,185.84){\rule{2.168pt}{0.800pt}}
\multiput(702.00,186.34)(4.500,-1.000){2}{\rule{1.084pt}{0.800pt}}
\put(711,184.34){\rule{1.927pt}{0.800pt}}
\multiput(711.00,185.34)(4.000,-2.000){2}{\rule{0.964pt}{0.800pt}}
\put(719,182.84){\rule{1.927pt}{0.800pt}}
\multiput(719.00,183.34)(4.000,-1.000){2}{\rule{0.964pt}{0.800pt}}
\put(727,181.84){\rule{1.927pt}{0.800pt}}
\multiput(727.00,182.34)(4.000,-1.000){2}{\rule{0.964pt}{0.800pt}}
\put(735,180.84){\rule{2.168pt}{0.800pt}}
\multiput(735.00,181.34)(4.500,-1.000){2}{\rule{1.084pt}{0.800pt}}
\put(744,179.84){\rule{1.927pt}{0.800pt}}
\multiput(744.00,180.34)(4.000,-1.000){2}{\rule{0.964pt}{0.800pt}}
\put(752,178.84){\rule{1.927pt}{0.800pt}}
\multiput(752.00,179.34)(4.000,-1.000){2}{\rule{0.964pt}{0.800pt}}
\put(760,177.84){\rule{1.927pt}{0.800pt}}
\multiput(760.00,178.34)(4.000,-1.000){2}{\rule{0.964pt}{0.800pt}}
\put(768,176.84){\rule{2.168pt}{0.800pt}}
\multiput(768.00,177.34)(4.500,-1.000){2}{\rule{1.084pt}{0.800pt}}
\put(777,175.84){\rule{1.927pt}{0.800pt}}
\multiput(777.00,176.34)(4.000,-1.000){2}{\rule{0.964pt}{0.800pt}}
\put(785,174.84){\rule{1.927pt}{0.800pt}}
\multiput(785.00,175.34)(4.000,-1.000){2}{\rule{0.964pt}{0.800pt}}
\put(793,173.84){\rule{2.168pt}{0.800pt}}
\multiput(793.00,174.34)(4.500,-1.000){2}{\rule{1.084pt}{0.800pt}}
\put(802,172.84){\rule{1.927pt}{0.800pt}}
\multiput(802.00,173.34)(4.000,-1.000){2}{\rule{0.964pt}{0.800pt}}
\put(810,171.84){\rule{1.927pt}{0.800pt}}
\multiput(810.00,172.34)(4.000,-1.000){2}{\rule{0.964pt}{0.800pt}}
\put(818,170.84){\rule{1.927pt}{0.800pt}}
\multiput(818.00,171.34)(4.000,-1.000){2}{\rule{0.964pt}{0.800pt}}
\put(835,169.84){\rule{1.927pt}{0.800pt}}
\multiput(835.00,170.34)(4.000,-1.000){2}{\rule{0.964pt}{0.800pt}}
\put(843,168.84){\rule{1.927pt}{0.800pt}}
\multiput(843.00,169.34)(4.000,-1.000){2}{\rule{0.964pt}{0.800pt}}
\put(851,167.84){\rule{1.927pt}{0.800pt}}
\multiput(851.00,168.34)(4.000,-1.000){2}{\rule{0.964pt}{0.800pt}}
\put(826.0,172.0){\rule[-0.400pt]{2.168pt}{0.800pt}}
\put(868,166.84){\rule{1.927pt}{0.800pt}}
\multiput(868.00,167.34)(4.000,-1.000){2}{\rule{0.964pt}{0.800pt}}
\put(876,165.84){\rule{1.927pt}{0.800pt}}
\multiput(876.00,166.34)(4.000,-1.000){2}{\rule{0.964pt}{0.800pt}}
\put(884,164.84){\rule{2.168pt}{0.800pt}}
\multiput(884.00,165.34)(4.500,-1.000){2}{\rule{1.084pt}{0.800pt}}
\put(859.0,169.0){\rule[-0.400pt]{2.168pt}{0.800pt}}
\put(901,163.84){\rule{1.927pt}{0.800pt}}
\multiput(901.00,164.34)(4.000,-1.000){2}{\rule{0.964pt}{0.800pt}}
\put(909,162.84){\rule{1.927pt}{0.800pt}}
\multiput(909.00,163.34)(4.000,-1.000){2}{\rule{0.964pt}{0.800pt}}
\put(893.0,166.0){\rule[-0.400pt]{1.927pt}{0.800pt}}
\put(926,161.84){\rule{1.927pt}{0.800pt}}
\multiput(926.00,162.34)(4.000,-1.000){2}{\rule{0.964pt}{0.800pt}}
\put(934,160.84){\rule{1.927pt}{0.800pt}}
\multiput(934.00,161.34)(4.000,-1.000){2}{\rule{0.964pt}{0.800pt}}
\put(917.0,164.0){\rule[-0.400pt]{2.168pt}{0.800pt}}
\put(950,159.84){\rule{2.168pt}{0.800pt}}
\multiput(950.00,160.34)(4.500,-1.000){2}{\rule{1.084pt}{0.800pt}}
\put(942.0,162.0){\rule[-0.400pt]{1.927pt}{0.800pt}}
\put(967,158.84){\rule{1.927pt}{0.800pt}}
\multiput(967.00,159.34)(4.000,-1.000){2}{\rule{0.964pt}{0.800pt}}
\put(959.0,161.0){\rule[-0.400pt]{1.927pt}{0.800pt}}
\put(984,157.84){\rule{1.927pt}{0.800pt}}
\multiput(984.00,158.34)(4.000,-1.000){2}{\rule{0.964pt}{0.800pt}}
\put(992,156.84){\rule{1.927pt}{0.800pt}}
\multiput(992.00,157.34)(4.000,-1.000){2}{\rule{0.964pt}{0.800pt}}
\put(975.0,160.0){\rule[-0.400pt]{2.168pt}{0.800pt}}
\put(1008,155.84){\rule{2.168pt}{0.800pt}}
\multiput(1008.00,156.34)(4.500,-1.000){2}{\rule{1.084pt}{0.800pt}}
\put(1000.0,158.0){\rule[-0.400pt]{1.927pt}{0.800pt}}
\put(1025,154.84){\rule{1.927pt}{0.800pt}}
\multiput(1025.00,155.34)(4.000,-1.000){2}{\rule{0.964pt}{0.800pt}}
\put(1017.0,157.0){\rule[-0.400pt]{1.927pt}{0.800pt}}
\put(1041,153.84){\rule{2.168pt}{0.800pt}}
\multiput(1041.00,154.34)(4.500,-1.000){2}{\rule{1.084pt}{0.800pt}}
\put(1033.0,156.0){\rule[-0.400pt]{1.927pt}{0.800pt}}
\put(1058,152.84){\rule{1.927pt}{0.800pt}}
\multiput(1058.00,153.34)(4.000,-1.000){2}{\rule{0.964pt}{0.800pt}}
\put(1050.0,155.0){\rule[-0.400pt]{1.927pt}{0.800pt}}
\put(1083,151.84){\rule{1.927pt}{0.800pt}}
\multiput(1083.00,152.34)(4.000,-1.000){2}{\rule{0.964pt}{0.800pt}}
\put(1066.0,154.0){\rule[-0.400pt]{4.095pt}{0.800pt}}
\put(1099,150.84){\rule{2.168pt}{0.800pt}}
\multiput(1099.00,151.34)(4.500,-1.000){2}{\rule{1.084pt}{0.800pt}}
\put(1091.0,153.0){\rule[-0.400pt]{1.927pt}{0.800pt}}
\put(1116,149.84){\rule{1.927pt}{0.800pt}}
\multiput(1116.00,150.34)(4.000,-1.000){2}{\rule{0.964pt}{0.800pt}}
\put(1108.0,152.0){\rule[-0.400pt]{1.927pt}{0.800pt}}
\put(1141,148.84){\rule{1.927pt}{0.800pt}}
\multiput(1141.00,149.34)(4.000,-1.000){2}{\rule{0.964pt}{0.800pt}}
\put(1124.0,151.0){\rule[-0.400pt]{4.095pt}{0.800pt}}
\sbox{\plotpoint}{\rule[-0.500pt]{1.000pt}{2.000pt}}%
\put(523.49,241.92){\usebox{\plotpoint}}
\put(531.09,261.23){\usebox{\plotpoint}}
\put(538.97,280.43){\usebox{\plotpoint}}
\put(547.64,299.28){\usebox{\plotpoint}}
\put(557.51,317.52){\usebox{\plotpoint}}
\put(568.31,335.25){\usebox{\plotpoint}}
\put(579.83,352.51){\usebox{\plotpoint}}
\put(592.86,368.63){\usebox{\plotpoint}}
\put(606.67,384.13){\usebox{\plotpoint}}
\put(621.65,398.44){\usebox{\plotpoint}}
\put(637.27,412.11){\usebox{\plotpoint}}
\put(653.84,424.53){\usebox{\plotpoint}}
\put(670.97,436.23){\usebox{\plotpoint}}
\put(689.21,446.01){\usebox{\plotpoint}}
\put(707.64,455.50){\usebox{\plotpoint}}
\put(726.61,463.85){\usebox{\plotpoint}}
\put(745.79,471.67){\usebox{\plotpoint}}
\put(765.22,478.96){\usebox{\plotpoint}}
\sbox{\plotpoint}{\rule[-0.600pt]{1.200pt}{1.200pt}}%
\put(441.01,753){\rule{1.200pt}{25.535pt}}
\multiput(438.51,806.00)(5.000,-53.000){2}{\rule{1.200pt}{12.768pt}}
\multiput(448.24,675.79)(0.503,-8.958){6}{\rule{0.121pt}{18.600pt}}
\multiput(443.51,714.39)(8.000,-83.395){2}{\rule{1.200pt}{9.300pt}}
\multiput(456.24,571.85)(0.503,-6.769){6}{\rule{0.121pt}{14.250pt}}
\multiput(451.51,601.42)(8.000,-63.423){2}{\rule{1.200pt}{7.125pt}}
\multiput(464.24,495.80)(0.502,-4.542){8}{\rule{0.121pt}{10.167pt}}
\multiput(459.51,516.90)(9.000,-52.899){2}{\rule{1.200pt}{5.083pt}}
\multiput(473.24,426.02)(0.503,-4.203){6}{\rule{0.121pt}{9.150pt}}
\multiput(468.51,445.01)(8.000,-40.009){2}{\rule{1.200pt}{4.575pt}}
\multiput(481.24,373.87)(0.503,-3.373){6}{\rule{0.121pt}{7.500pt}}
\multiput(476.51,389.43)(8.000,-32.433){2}{\rule{1.200pt}{3.750pt}}
\multiput(489.24,330.23)(0.503,-2.845){6}{\rule{0.121pt}{6.450pt}}
\multiput(484.51,343.61)(8.000,-27.613){2}{\rule{1.200pt}{3.225pt}}
\multiput(497.24,295.38)(0.502,-2.055){8}{\rule{0.121pt}{4.967pt}}
\multiput(492.51,305.69)(9.000,-24.691){2}{\rule{1.200pt}{2.483pt}}
\multiput(506.24,261.07)(0.503,-2.015){6}{\rule{0.121pt}{4.800pt}}
\multiput(501.51,271.04)(8.000,-20.037){2}{\rule{1.200pt}{2.400pt}}
\multiput(514.24,233.57)(0.503,-1.713){6}{\rule{0.121pt}{4.200pt}}
\multiput(509.51,242.28)(8.000,-17.283){2}{\rule{1.200pt}{2.100pt}}
\end{picture}
\caption{\label{b2} Ground states in $c^a~-~c^b$ parameter plane for $0< c^{ab} \leq 2\gamma B$ and $B >0$, with $N_a=N_b=N $.  A state which is disentangled is
written in the form of $|S_a^m,S_a^m\rangle_a|S_b^m,S_b^m\rangle_b$.   A state which may be entangled is written in the general form of $|S^m_a,S^m_b,S^m,S^m\rangle$.
The borders of $|N,N,n,n\rangle$ in the lower left part is $c^a = c^{ab}$ and $ c^b = c^{ab}$. The ground state here is $|N_a,N_b,n,n\rangle$ for generic $N_a$ and $N_b$, which is entangled unless $n=N_a+N_b$. The complete hyperbola is $c^ac^b = (c^{ab})^2$, above which all the ground states are continuous connected, forming a single quantum phase. It is $|0,0\rangle_a|0,0\rangle_b$ in the upper right part $c^a \geq  2\gamma B$ and $ c^b \geq 2\gamma B$. Below it the ground state is $|0,0\rangle_a  |n_b,n_b\rangle_b$, which is bounded on the left by  $2\gamma B \leq \frac{c^{a}-c^{ab}}{c^{b}-c^{ab}}c^{b}$, and on the bottom by $2\gamma B \leq \frac{c^{a}-c^{ab}}{c^{ab}-c^b}c^{b}$. Surrounded by the former boundary and $c^{a} = c^{b}$ is the regime  numbered as B2b, where  the ground state is $|n_{a3},n_{a3}\rangle_a |n_{b1},n_{b1}\rangle_b$. In the regime numbered as C2, i.e. between $2\gamma B \leq \frac{c^{a}-c^{ab}}{c^{ab}-c^b}c^{b}$ and $c^{a}c^{b} = {(c^{ab})^2}$, the ground state is  $|n_{a3},n_{b2},n_{b2}-n_{a3},n_{b2}-n_{a3}\rangle$, which is entangled except on the boundary with $|0,0\rangle_a  |n_b,n_b\rangle_b$.
Similarly, the ground state is $ |n_a,n_a\rangle_a |0,0\rangle_b $, in the regime bounded by  $c^a =2\gamma B$ on the right, $2\gamma B =  {\frac{c^{b}-c^{ab}}{c^{a}-c^{ab}}}c^{a}$ below  and $2\gamma B = {\frac{c^{b}-c^{ab}}{c^{ab}-c^a}}c^{a}$ on the left.  In the regime numbered as D2b, as surrounded by
$c^{a} = c^{b}$ and $2\gamma B= \frac{c^{b}-c^{ab}}{c^{a}-c^{ab}}c^{a}$, the ground state is $|n_{a1},n_{a1}\rangle_a|n_{b3},n_{b3}\rangle_b$. In the regime numbered as E2, i.e. between $c^{a}c^{b} = {(c^{ab})^2}$  and $2\gamma B=\frac{c^{b}-c^{ab}}{c^{ab}-c^{a}}c^{a}$, the ground state is  $|n_{a2},n_{b3},n_{a2}-n_{b3},n_{a2}-n_{b3}\rangle$, which is entangled except on the boundary with  $ |n_a,n_a\rangle_a |0,0\rangle_b $. For the regimes $c^a > c^{ab} >c^b$ and $c^b> c^{ab} >c^a$,  the results are subject to the condition $N>N^*$. $n$, $n_{a}$, $n_{b}$, $n_{a2}$, $n_{b2}$  $n_{a3}$, $n_{b3}$  and $N^*$ are defined in the main text and  Table~\ref{table}. }
\end{figure*}
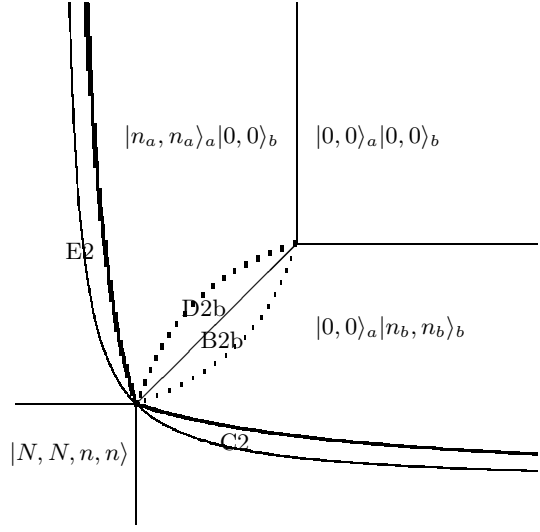

We have also  determined the ground states in the regime $0<c^{ab} \leq 2\gamma B$ and $c^ac^b > (c^{ab})^2$. This regime is divided into seven  subregimes, but the ground states are all continuously connected on the boundaries between these subregimes, as depicted in $c^a-c^b$ phase diagram for a given value of $c^{ab}$ with $0 < c^{ab} < 2\gamma B$ (FIG.~\ref{b2}), drawn according to Table~\ref{table}.

For $c^b > 2\gamma B$ and $c^a > 2\gamma B$, the ground state is $|0,0,0,0\rangle = |0,0\rangle_a |0,0\rangle_b$.  In the regime defined by $c^b\leq 2\gamma B$,  $2\gamma B \leq \frac{c^a-c^{ab}}{c^b-c^{ab}} c^b$ (i.e. below the hyperbola $2\gamma B = \frac{c^a-c^{ab}}{c^b-c^{ab}} c^b$), and $2\gamma B \leq \frac{c^a-c^{ab}}{c^{ab}-c^{b}} c^b$ (i.e. above the hyperbola $2\gamma B = \frac{c^a-c^{ab}}{c^{ab}-c^{b}} c^b$), the ground state is $|0,0\rangle_a|n_b,n_b\rangle_b$, where $n_b \equiv Int[\frac{\gamma
B}{c^{b}}-\frac{1}{2}]$, which reduces to $0$ when $c^b=2\gamma B$.
In the regime defined by $c^{a} \geq c^{b}>c^{ab}$ and  $2\gamma B>\frac{c^{a}-c^{ab}}{c^{b}-c^{ab}}c^{b}$, that is, surrounded by $c^a=c^b$ and $2\gamma B =\frac{c^{a}-c^{ab}}{c^{b}-c^{ab}}c^{b}$, the ground state is
$|n_{a3},n_{a3}\rangle_a |n_{b1},n_{b1}\rangle_b$, where $n_{a3} \equiv Int[\frac{2\gamma
B|c^{b}-c^{ab}|-c^{b}(c^{a}-c^{ab})}{2[c^{a}c^{b}-(c^{ab})^2]}]$,   $n_{b1}\equiv Int[\frac{\gamma
B-c^{ab}n_{a3}}{c^{b}}-\frac{1}{2}]$.
These states are all disentangled.
The ground state is $|n_{a3},n_{b2},n_{b2}-n_{a3},n_{b2}-n_{a3}\rangle$, where
$n_{b2} \equiv Int[\frac{\gamma
B+c^{ab}n_{a3}}{c^{b}}-\frac{1}{2}]$, in the regime numbered as C2 and
defined by $c^a>c^{ab} > c^b$,
$c^{a}c^{b} > {(c^{ab})^2}$ and $2\gamma B \geq \frac{c^{a}-c^{ab}}{c^{ab}-c^{b}}c^{b}$, for $N_a=N_b=N >N^*$, where $N^*$ is given in (\ref{nn}).  This state is always entangled except on the boundary  $2\gamma B=\frac{c^{a}-c^{ab}}{c^{ab}-c^{b}}c^{b}$, where the state reduces to $|0,0\rangle_a|n_b,n_b\rangle_b$.
By substituting the boundary values of $c^a$ and $c^b$ to the values of $S_a$, $S_b$ and $S$ that depend on them, it is not difficult to see that the ground states in each subregime continuously connected with those in its neighboring subregimes.

Similarly, by exchanging the labels $a$ and $b$, we know the ground states in the part of $c^ac^b > (c^{ab})^2$ with $c^a \leq c^b$  and $c^a \leq 2\gamma B$.  In the subregime defined by $c^a\leq 2\gamma B$, $2\gamma B \leq \frac{c^b-c^{ab}}{c^a-c^{ab}} c^a$ (i.e. above the hyperbola $2\gamma B = \frac{c^b-c^{ab}}{c^a-c^{ab}} c^a$), and $2\gamma B \leq \frac{c^b-c^{ab}}{c^{ab}-c^{a}} c^a$ (i.e. on the right of the hyperbola $2\gamma B = \frac{c^b-c^{ab}}{c^{ab}-c^{a}} c^a$), the ground state is $|n_a,n_a\rangle_a|0,0\rangle_b$, where $n_a \equiv Int[\frac{\gamma
B}{c^{a}}-\frac{1}{2}]$.  In the regime defined by $c^{b} \geq c^{a} > c^{ab}$ and  $2\gamma B > \frac{c^{b}-c^{ab}}{c^{a}-c^{ab}}c^{a}$, that is, surrounded by $c^a=c^b$ and $2\gamma B=\frac{c^{b}-c^{ab}}{c^{a}-c^{ab}}c^{a}$, the ground state is
$|n_{a1},n_{a1}\rangle_a|n_{b3},n_{b3}\rangle_b $, where $n_{b3} \equiv Int[\frac{2\gamma
B|c^{a}-c^{ab}|-c^{a}(c^{b}-c^{ab})}{2[c^{a}c^{b}-(c^{ab})^2]}]$,  $n_{a1}\equiv Int[\frac{\gamma
B-c^{ab}n_{b3}}{c^{a}}-\frac{1}{2}]$.
These states are all disentangled.
Finally, the ground state is $|n_{a2},n_b,n_{a2}-n_b,n_{a2}-n_b\rangle$, where
$n_{a2} \equiv Int[\frac{\gamma
B+c^{ab}n_b}{c^{b}}-\frac{1}{2}]$, in the regime numbered as E2 and
defined by $c^b>c^{ab} > c^a$,
$c^{a}c^{b} > {(c^{ab})^2}$ and $2\gamma B \geq \frac{c^{b}-c^{ab}}{c^{ab}-c^{a}}c^{a}$, for $N_a=N_b=N >N^*$.
This state is always entangled except on the boundary  $2\gamma B = \frac{c^b-c^{ab}}{c^{ab}-c^{a}} c^a$, where the state reduces to $|n_a,n_a\rangle_a|0,0\rangle_b$.
The ground states in each subregime continuously connected with those in its neighboring subregimes.

Also, the ground states $|n_{a1},n_{a1}\rangle_a|n_{b3},n_{b3}\rangle_b $ and $|n_{a3},n_{a3}\rangle_a|n_{b1},n_{b1}\rangle_b $ are continuously connected on the boundary $c^{ab}<c^a=c^b\leq 2\gamma B$, with $n_{a3}=n_{b1}=n_{a1}=n_{b3}=\frac{2\gamma B- c^a}{2(c^a+c^{ab})}$.

But at the point $c^a=c^b=c^{ab}$, the ground states in the six subregimes of $c^ac^b > (c^{ab})^2$ converging at this point are discontinuous  with other, as one can see from $n_a=n_b=Int(\frac{\gamma B}{c^{ab}}-\frac{1}{2})$, $n_{a3}=n_{b1}=n_{b3}=n_{a1}=Int(\frac{\gamma B}{2c^{ab}}-\frac{1}{4})$ and
$n_{b2}=n_{a2}= Int(\frac{3\gamma B}{2c^{ab}}-\frac{3}{4})$ at this point.

As $c^{ab} \rightarrow 2\gamma B$, $c^a=c^b=2\gamma B$ approaches $c^a=c^b=c^{ab}$, hence the regimes B2b and D2b tend to vanish.

\section{$c^{ab} \geq 2\gamma B$ \label{anti2}}

For $c^{ab} \geq 2\gamma B$~\cite{shi4}, it has been known that the ground state is  $|0,0\rangle_a|0,0\rangle_b$ for  $c^{a} > c^{ab}$ and  $c^{b} > c^{ab}$,  and is  $|N,N,0,0\rangle $ for  $c^{a} < c^{ab}$, $c^{b} < c^{ab}$ and  $N_a=N_b=N$. This result is consistent with with those obtained for $0 < c^{ab} \leq 2\gamma B$. Hence for the regime $c^{a} > c^{ab}$ and  $c^{b} > c^{ab}$ and the regime   $c^{a} < c^{ab}$, $c^{b} < c^{ab}$,  $N_a=N_b=N$, the ground states for $c^{ab} \leq 2\gamma B$ and those for  $c^{ab} \geq 2\gamma B$ are continuously connected at $c^{ab} = 2\gamma B$.

Especially, in the cases we have studied, under the condition $N_a=N_b=N$, in the regime $c^a <c^{ab}$ and $c^b <c^{ab}$ and in the regime $c^{a} > 2\gamma B$ and $c^{b} > 2\gamma B$,  $c^{ab} = 2\gamma B >0 $ is not a boundary, i.e. the ground states are respectively the same in these two regimes for $c^{ab} \geq 2\gamma B$ and for $0 < c^{ab} \leq 2\gamma B$.

\begin{figure*}
\setlength{\unitlength}{0.240900pt}
\ifx\plotpoint\undefined\newsavebox{\plotpoint}\fi
\begin{picture}(1500,900)(0,0)
\sbox{\plotpoint}{\rule[-0.200pt]{0.400pt}{0.400pt}}%
\put(530,245){\makebox(0,0)[l]{$|N,N,0,0\rangle$}}
\put(816,654){\makebox(0,0)[l]{$|0,0\rangle_a|0,0
\rangle_{b}$}}
\put(330,450){\line(1,0){820}}
\put(740,40){\line(0,1){819}}
\put(330,450){\usebox{\plotpoint}}
\put(330.0,450.0){\rule[-0.200pt]{197.297pt}{0.400pt}}
\end{picture}
\caption{ Ground states in $c^a~-~c^b$ parameter plane for $c^{ab} \geq 2\gamma B$ and $B >0$, with $N_a=N_b=N$. The ground state is the entangled  state $|N,N,0,0\rangle$ in the regime $c^a < c^{ab}$ and $ c^b < c^{ab}$. It is the disentangled  state  $|0,0\rangle_a|0,0\rangle_b$ in the regime $c^a >  c^{ab}$ and $ c^b >  c^{ab}$. }

\end{figure*}
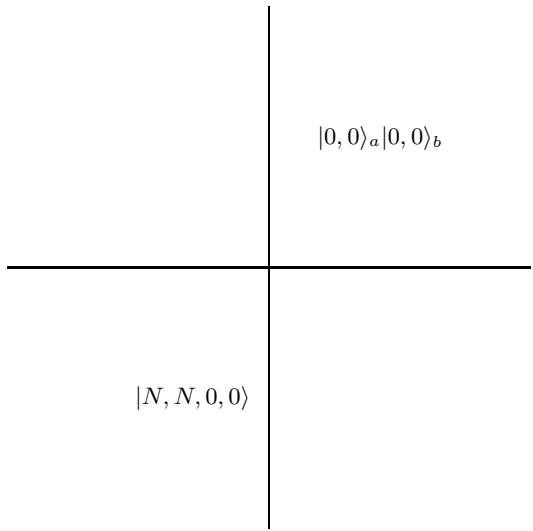

\section{Quantum phase transitions \label{qpt} }

\subsection{Quantum phase transitions at $c^{a}=c^{b}=c^{ab}$}

In the regime of $0<c^{ab}<2\gamma B$, quantum phase transitions take place at $c^{a}=c^{b}=c^{ab}$, which is the boundary between the two phases discussed above for $0<c^{ab} \leq 2\gamma B$. It is a point on $c^a-c^b$ plane with given $c^{ab}$ and $B$, and is a line in the three dimensional $c^a-c^b-c^{ab}$ subspace with a given $B$, and is a two-dimensional surface in the four dimensional $c^a-c^b-c^{ab}-B$ space.

At $c^{a}=c^{b}=c^{ab}$, any state in the form of  $|S_a,S_b,n,n\rangle$ with arbitrary legitimate values of $S_a$, $S_b$ and $n$ is a ground state. Therefore its degenerate ground state space includes the  ground states in all the seven regimes  we have studied that contact at this degenerate point, that is, the ground states of the six subregimes of $c^ac^b > (c^{ab})^2$ neighboring at $c^a=c^b=c^{ab}$,  as well as the ground state $|N_a,N_b,n,n\rangle$ in the regime $c^a <c^{ab}$ and $c^b < c^{ab}$. Therefore in entering  $c^{a}=c^{b}=c^{ab}$ from one of the two phases, the ground state remains as the original, and then discontinues in entering the any of the other six regimes. Note that there is also a discontinuity in transiting, through the critical point $c^a=c^b=c^{ab}$,  from one of the six regimes belonging to the same phase.

In any of these seven regimes converging at the point  $c^{a}=c^{b}=c^{ab}$, we always have $S^m=S_z^m=n$. Therefore, the quantum phase transition is a continuous transition.

If $B \rightarrow 0$, for $0 < c^{ab} < 2\gamma B$, the range of $c^{ab}$ has to be diminished as well, hence in this regime  $c^{ab} \rightarrow 0$ too. Consequently the regime of $|0,0\rangle_a|0,0\rangle_b$ expands to occupy the whole first quadrant of $c^a-c^b$ plane, while the other regimes in the first quadrant have to be diminished. On the other hand, we have $n \rightarrow 0$ as a consequence of $B \rightarrow 0$. Therefore the ground states approach to the corresponding ones for $c^{ab} >2\gamma B$, but $c^{ab}$ being  infinitesimally positive is  qualitatively different from the case of $c^{ab}=0$, as will be discussed in the next subsection.

For $c^{ab} \geq 2\gamma B$, the quantum phase transition from the ground state $|N,N,0,0\rangle$ in the regime $c^a < c^{ab}$ while  $c^b < c^{ab}$ to the ground state $|0,0\rangle_a|0,0\rangle_b$ in the regime $c^a > c^{ab}$ and  $c^b >  c^{ab}$ is similar to the one for  $c^{ab} < 2\gamma B$, with $n$ in the latter becoming $0$. Hence it is also a continuous   quantum phase transition.

\subsection{Quantum phase transitions from $c^{ab}=0$ to $c^{ab} >0$ }

\begin{figure*}
\setlength{\unitlength}{0.240900pt}
\ifx\plotpoint\undefined\newsavebox{\plotpoint}\fi
\begin{picture}(1500,900)(0,0)
\sbox{\plotpoint}{\rule[-0.200pt]{0.400pt}{0.400pt}}%
\put(330,274){\makebox(0,0)[l]{$|N_a,N_b,n,n
\rangle$}}
\put(668,664){\makebox(0,0)[l]{$|n_a,n_a
\rangle_a|0,0
\rangle_{b}$}}
\put(668,469){\makebox(0,0)[l]{$|n_a,n_a
\rangle_a|n_b,n_b
\rangle_{b}$}}
\put(993,664){\makebox(0,0)[l]{$|0,0
\rangle_a|0,0
\rangle_{b}$}}
\put(993,664){\makebox(0,0)[l]{$|0,0
\rangle_a|0,0
\rangle_{b}$}}
\put(993,469){\makebox(0,0)[l]{$|0,0
\rangle_a|n_b,n_b
\rangle_{b}$}}
\put(330,373){\line(1,0){902}}
\put(662,606){\line(1,0){570}}
\put(662,40){\line(0,1){819}}
\put(970,373){\line(0,1){486}}
\end{picture}

\caption{\label{b0+} Ground states in $c^a-c^b$ parameter plane for $c^{ab} = 0+$ and $B>0$. The two horizontal lines are  $c^a = 0$ and   $c^a = 2\gamma B$, while the two vertical lines are   $c^b = 0$ and   $c^b = 2\gamma B$. The ground states are in the disentangled form of $|S_a^m,S_a^m\rangle_a|S_b^m,S_b^m\rangle_b$ in all the regimes with $c^a >0$ and  $c^b > 0$, but are in the form of  $|S_a^m,S_b^m,S^m,S^m\rangle$ in the regime with
$c^a <  0$ and  $c^b< 0$. }
\end{figure*}
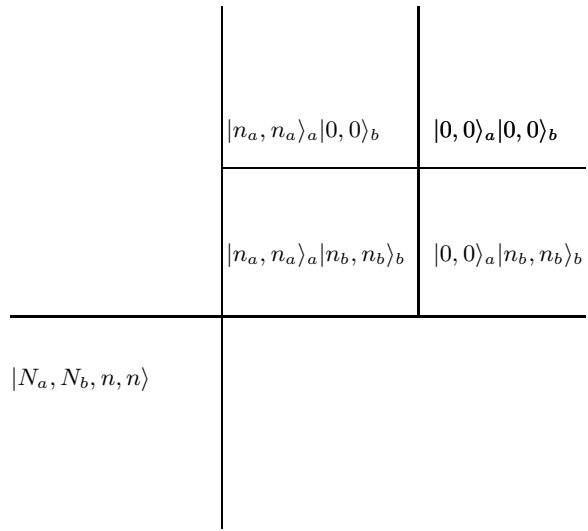

On the other hand, when $c^{ab}\rightarrow 0$ from the positive side, i.e, $c^{ab}=0+$, under a given $B$, the regime of $|0,0\rangle_a|0,0\rangle_b$ remains unchanged, while the regime of $|N_a,N_b,n,n\rangle$ approaches the third quadrant, and the hyperbola $c^a c^b =(c^{ab})^2$ approaches the positive $c^a$ and $c^b$ axes. The regime of $|0,0\rangle_a|n_b,n_b\rangle_b$ approaches $c^a \geq 2\gamma B$ while $0 < c^b < 2\gamma B$, similarly the regime of $|n_a,n_a\rangle_a|0,0\rangle_b$ approaches $c^b \geq 2\gamma B$ while $0 < c^a < 2\gamma B$. The regimes B2b plus D2b become the regime $0 <c^a< 2\gamma B$ and  $0 <c^b< 2\gamma B$, where the ground state approaches $|n_a,n_a\rangle_a|n_b,n_b\rangle_b$. Note that the subregime of $c^ac^b > (c^{ab})^2$ with  $c^a > c^{ab} > c^b$ and that with $c^b > c^{ab} > c^a$, including the two subregimes of entangled ground states, tend to vanish. In FIG.~\ref{b0+}, we draw the $c^a-c^b$ phase diagram for a given $B$ while $c^{ab}=0+$, referring to that $c^{ab} \rightarrow 0$ from positive.

Comparing the ground states  of $c^{ab}=0$ (FIG.~\ref{b0}) and $c^{ab}=0+$ (FIG.~\ref{b0+}), we can see  there are discontinuities between $c^{ab} >0$ and  $c^{ab} = 0$.  First, in the third quadrant, the ground state is $|N_a,N_a\rangle_a|N_b,N_b\rangle_b$ for $c^{ab} = 0$, discontinuing with $|N_a,N_b,n,n\rangle$ for  $c^{ab} > 0$. This  discontinuity already exists when $B=0$~\cite{shi5}. This quantum phase transition is first order as $S^m$ has a discontinuity except in the special case $n=N_a+N_b$, for which the transition becomes continuous.

Moreover, there are also other discontinuities, which are induced by $B>0$.
On a $c^a-c^b$ plane, the boundaries $c^a = 2\gamma B$ and $c^b=2\gamma B$ exist both for $c^{ab}=0+$ and  $c^{ab} =0$. However, the other two boundaries are different, that is, they are $c^a =0$ and $c^b =0$ for $c^{ab} = 0+$, but are $c^a=\frac{2\gamma B}{2N_a+1}$ and  $c^b = \frac{2\gamma B}{2N_b+1}$ for $c^{ab} = 0$, though the differences diminish as $N_a$ and $N_b$ approach infinities.

Consequently, there are five discontinuities induced by $B$ for $c^a >0$ and $c^b >0$. In the regime $0 \leq c^a \leq \frac{2\gamma B}{2N_a+1}$ and  $c^b \geq 2\gamma B$,  the ground state discontinues from $|n_a,n_a\rangle_a|N_b,N_b\rangle$ for $c^{ab} = 0+$  to $|N_a,N_a\rangle|N_b,N_b\rangle$ for  $c^{ab} = 0$.  This is a first order quantum phase transition except in the special case of $n_a=N_a$, in which the transition becomes continuous.

In the regime $0 \leq c^a \leq \frac{2\gamma B}{2N_a+1}$ and $\frac{2\gamma B}{2N_b+1} \leq c^b \leq 2\gamma B $,  the ground state discontinues from $|n_a,n_a\rangle_a|n_b,n_b\rangle$ for $c^{ab} = 0+$  to $|N_a,N_a\rangle|n_b,n_b\rangle$ for $c^{ab} = 0$.  This is a first order quantum phase transition except in the special case of $n_a=N_a$.

In the regime $0 \leq c^a \leq \frac{2\gamma B}{2N_a+1}$ and $0 \leq c^b \leq \frac{2\gamma B}{2N_b+1}$,  the ground state discontinues from $|n_a,n_a\rangle_a|n_b,n_b\rangle$ for $c^{ab} = 0+$  to $|N_a,N_a\rangle_a|N_b,N_b\rangle$ for $c^{ab} = 0$.  This is a first order quantum phase transition except in the special case of $n_a=N_a$ while $n_b=N_b$.

In the regime $ \frac{2\gamma B}{2N_a+1}  \leq c^a \leq 2\gamma B$ and $0 \leq c^b \leq \frac{2\gamma B}{2N_b+1}$,  the ground state discontinues from $|n_a,n_a\rangle_a|n_b,n_b\rangle$ for $c^{ab} = 0+$  to $|n_a,n_a\rangle_a|N_b,N_b\rangle$ for $c^{ab} = 0$.  This is a first order quantum phase transition except in the special case of $n_b=N_b$.

In the regime $ c^a \geq 2\gamma B$ and $0 \leq c^b \leq \frac{2\gamma B}{2N_b+1}$,  the ground state discontinues from $|0,0\rangle_a|n_b,n_b\rangle$ for $c^{ab} = 0+$  to $|0,0\rangle_a|N_b,N_b\rangle$ for $c^{ab} = 0$.  This is a first order quantum phase transition except in the special case of $n_b=N_b$.

Therefore, we find five places of quantum phase transitions from $B=0$ to $B >0$. In other words,  the entire subspace of $B=0$ is critical.

\section{Interspecies entanglement \label{entanglement} }

Our results indicate that a necessary condition for the ground state to be entangled between the two species is $c^{ab} >0$. We have found that the ground state is an entangled state entangled $|N_a, N_b,n,n\rangle$ for $0 < c^{ab} < 2\gamma B$, $c^a<c^{ab}$ and $c^{b}<c^{ab}$. In case $N_a=N_B=N$, we have also found that the ground state is a maximal entangled state  $|N, N, 0,0\rangle$ for $c^{ab} > 2\gamma B$, $c^a<c^{ab}$ and $c^{b}<c^{ab}$.

With interspecies entanglement, the occupation number of each spin state of each species is subject to fluctuation~\cite{shi1}.
However,  even in absence of interspecies entanglement, such fluctuations can still exist, and there can be occupation number entanglement among different single particle states defined by the spin and the species. Such is the singlet ground state of single species of spinor atoms, for example.  Therefore particle number fluctuations are not satisfactory characterizations of interspecies entanglement caused by interspecies spin exchanges.

A better characterization  is an interspecies correlation function, e.g. $\langle N_{a\sigma}N_{b\sigma'}\rangle - \langle N_{a\sigma}\rangle \langle N_{b\sigma'}\rangle$, which vanishes for disentangled state and is nonvanishing if there is interspecies entanglement~\cite{shi1}.

One can also simply use the  spin of freedom of the two species to discuss the entanglement between the two species, treating the two species like two giant spins. Then, of course, the entanglement entropy can be calculated. For state $|S^m_a,S^m_b,S^m,S^m\rangle$, the entanglement entropy is
\begin{equation}
{\cal E} = -\sum_{S_{bz}=-S^m_b}^{S_b^m} |g(S_{bz})|^2 \log_{2S^m_b+1}  |g(S_{bz} )|^2,
\end{equation}
where  it is assumed that $S^m_a \geq S^m_b$, $g(S_{bz}) \equiv \langle S^m_a,S^m-S_{bz}; S^m_b, S_{bz}|S^m_a,S^m_b,S^m,S^m\rangle $ is the Clebsch-Gordan coefficient. If $S^m_a \leq S^m_b$, the subscripts $a$ and $b$ are exchanged. ${\cal E}=0$ for disentangled states, while ${\cal E} =1$ for state $|N,N,0,0\rangle$.

We also note that there is a simple yet experimentally measurable quantity as a characterization of the interspecies entanglement. This is just the total magnetization $S^m$. If  $S^m=S_a^m+S_b^m$, there can only be one term in the Schmidt decomposition of the ground state in terms of $|S^m_a,S_{az}\rangle$ and $|S_b^m,S_{bz}\rangle$, consequently it is disentangled. If $S^m=|S_a^m-S_b^m|$, the ground state is entangled, as there is $2L+1$ terms in the Schmidt decomposition, where $L$ represents represents the smaller one of $S_a^m$ and $S_b^m$.

\section{summary \label{summary} }

We have obtained most of the ground states of a mixture of spin-1 Bose gases with interspecies  spin coupling in presence of a magnetic field. For $c^{ab} \leq 0$,
the ground states, which are all disentangled, belong to a single quantum phase. For $c^{ab} <0$, a magnetic field modifies the ground states and the boundaries between them. For $c^{ab}=0$, a magnetic field induces some crossover regimes, hence discontinuities between ground states in different quadrants of $c^a-c^b$ plane in absence of a magnetic field now disappear.

For $c^a < c^{ab}$ and $c^b <c^{ab}$, a magnetic field divides the regime of $c^{ab} >0$ into two regimes continuously connecting at $c^{ab}=2\gamma B$. For $0 < c^{ab} \leq 2 \gamma B$, in this regime of $c^a$ and $c^b$, the ground state is $|N_a,N_b,n,n\rangle$, where  $n \equiv Int(\frac{\gamma B
}{c^{ab} }- \frac{1}{2})$ satisfies $|N_a-N_b| \leq n \leq N_a+N_b$. For $N_a=N_b$, it is continuously connected with  $|N,N,0,0\rangle$ for $c^{ab} \geq 2\gamma B$ in the same ranges of $c^a$ and $c^b$. It is discontinuous with $|N_a,N_a\rangle_a|N_b,N_b\rangle_b$ for $c^{ab}=0$ in the same ranges of  $c^a$ and  $c^b$, as in the case without a magnetic field. This is a first order quantum phase transition except $n=N_a+N_b$.

Moreover, a magnetic field causes discontinuities of ground states between $c^{ab} >0$ and $c^{ab} =0$ in the first quadrant of $c^a-c^b$ plane. These discontinuities do not exist in absence of a magnetic field. As $c^{ab} \rightarrow 0$ but remains positive, a magnetic field causes the division of the first quadrant into four regimes with continuous connecting ground states, as shown in FIG.~\ref{b0+}, while for $c^{ab}=0$ there are nine regimes with continuous connecting ground states, as shown in FIG.~\ref{b0}.  The boundaries of the ground states in these two cases do not match, leading to discontinuities between $c^{ab} >0$ and $c^{ab} =0$. These are usually first  order quantum phase transitions except in  some special cases.

$c^{a}=c^{b}=c^{ab} >0$  is extremely interesting place, where continuous  quantum phase transitions take place  no matter whether there is a magnetic field and no matter what is the actual value.

In terms of bosonic degrees of freedom, the general expression and its composite structure of $|S_a,S_b,S,S\rangle$ have been discussed previously~\cite{shi4a,shi5}. It will be very appealing to study the different physical consequences and the experimental probes of the crossovers and the discontinuities or quantum phase transitions of the ground states, and the effects of interspecies entanglement.

\acknowledgments

This work was supported by the National Science Foundation of China (Grant No. 11074048)  and the Ministry of Science and Technology of China (Grant No. 2009CB929204).
\vspace{1cm}

{\it Note added}:  after this paper had been initially submitted to Phys. Rev. A on September 15 2010, there appeared a paper treating the subject in a mean field approach~\cite{xu}.

\appendix

\section{$S_a^m$, $S_b^m$ and $S^m$ for $c^{ab} < {0}$, $B > 0$ }

In this appendix, we find out $S_a^m,S_b^m$ and $S^m$, in which $E$ is minimal, in the case of $c^{ab} < {0}$ and $B > 0$.
In the discussions, $E$ always represent the energy as low as can be determined in the regime under discussion, i.e. the meaning of $E$ keeps updating.

With $c^{ab} < 0$, $E$ is minimal when $S=S_a+S_b$. Hence the ground state with $S_z=S$ is always disentangled. Now

\begin{equation}
E=\frac{c^{a}}{2}S_a(S_a+1)
+\frac{c^{b}}{2}S_b(S_b+1)+c^{ab}S_aS_b-\gamma B(S_a+S_b).
\end{equation}
Thus
\begin{equation}
\frac{\partial{E}}{\partial{S_a}}=
c^{a}S_a+c^{ab}S_b+\frac{c^{a}}{2}-\gamma B,
\end{equation}
\begin{equation}
\frac{\partial{E}}{\partial{S_b}}=
c^{b}S_b+c^{ab}S_a+\frac{c^{b}}{2}-\gamma B.
\end{equation}

We consider three subcases in the following.

\subsection{$c^{a}\leq 0$, $c^{b} \leq 0$ \label{casea1}}

In this subcase, $\frac{\partial{E}}{\partial{S_a}}<0$,
$\frac{\partial{E}}{\partial{S_b}}<0$, hence $S^m_a=N_a$,
$S^m_b=N_b, S^m=N_a+N_b$.

\subsection{$c^{a}\leq 0$, $c^{b}>0$}

In this subcase,
$\frac{\partial{E}}{\partial{S_a}}<0$, hence $S_a^m=N_a$,  \begin{equation}
E(N_a,S_b) =\frac{c^{b}}{2}S_b(S_b+1)+c^{ab}N_aS_b-\gamma BS_b+const.
\end{equation}

\begin{figure}
\begin{center}
\scalebox{0.6}{\includegraphics[126,538][424,750]{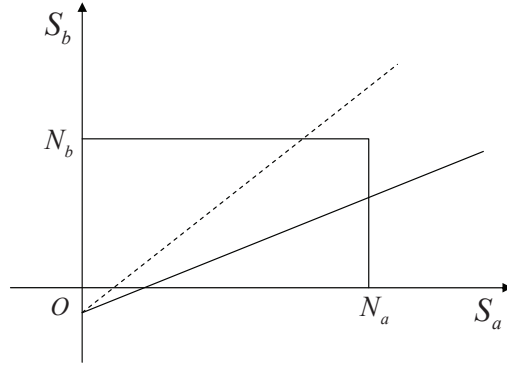}}
\end{center}
\caption{All possible values of $(S_a,S_b)$ are within the rectangular $0 \leq S_a \leq N_a$ and $0 \leq S_b \leq N_b$.
$\frac{\partial{E}}{\partial{S_b}}=0$ is represented as the dashed line in case $c^a \leq 0$ and $0 <c^{b} \leq \frac{2\gamma B-2N_ac^{ab}}{2N_b+1}$, and is represented as the solid line in case $c^a \leq 0$ and $ \frac{2\gamma B -2N_ac^{ab}}{2N_b+1} \leq c^{b} \leq 2\gamma B-2 N_a c^{ab}$.} \label{mixa}
\end{figure}

We represent all the values of $S_a$ and $S_b$ as points $(S_a, S_b)$ within the rectangular defined by $0 \leq S_a \leq N_a$ and  $0 \leq S_b \leq N_b$ on
$S_a$-$S_b$ plane (FIG.~\ref{mixa}).
$\frac{\partial{E}}{\partial{S_b}}=0$  defines a stationary line.
The points above this line satisfy $\frac{\partial{E}}{\partial{S_b}}>0$, while the points below the line satisfy  $\frac{\partial{E}}{\partial{S_b}}<0$.  One can see three possibilities.

\subsubsection{ $0 <c^{b} \leq \frac{2\gamma B-2N_ac^{ab}}{2N_b+1}$}

The stationary
line, depicted as the dashed line in FIG.~\ref{mixa}, crosses with the line $S_b=N_b$. Hence all points with $S_a=N_a$
satisfy $\frac{\partial{E}}{\partial{S_b}} \leq 0$. Consequently
$S_a^m=N_a$, $S_b^m=N_b$, $S^m=N_a+N_b$. Note that this regime so defined can be combined with   case \ref{casea1},  with the same result.

\subsubsection{  $ \frac{2\gamma B -2N_ac^{ab}}{2N_b+1} \leq c^{b} \leq 2\gamma B-2 N_a c^{ab}$}

The stationary line, depicted as the solid line in FIG.~\ref{mixa}, crosses with the line $S_a=N_a$. The crossing point gives the minimal energy.
Hence $S_a^m=N_a$, $S_b^m=n_1$, with
\begin{equation}
n_{1}' \equiv Int[\frac{\gamma B+N_a|c^{ab}}{c^{b}}-\frac{1}{2}],
\end{equation}
where $Int(x)$ represents the integer closest to $x$ and in the legitimate range  of $S_b$, i.e. now $0\leq Int(x) \leq N_b$.  $S^m=N_a + n_1$.

\subsubsection{ $c^{b} \geq 2\gamma B-2 N_a c^{ab}$}

 All points $(S_a, S_b)$ in the rectangular satisfy $\frac{\partial{E}}{\partial{S_b}} > 0$. Therefore $S_b^m =0$, $S_a^m=S^m =N_a$.

\subsection{$c^{a}>0$, $c^{b}\leq 0$}

One simply exchanges the subscripts or superscripts  $a$ and $b$ in the preceding subcase. Thus there are also three possibilities.

\subsubsection{ $0<c^{a} \leq \frac{2\gamma B-2N_b c^{ab}}{2N_a+1}$}

$S_a^m=N_a$, $S_b^m=N_b$, $S^m=N_a+N_b$. This regime can be combined with case \ref{casea1}, without the same result.

\subsubsection{ $ \frac{2\gamma B-2N_b c^{ab}}{2N_a+1} \leq c^{a} \leq  2\gamma B-2 N_b c^{ab}$.}

$S_a^m=n_{b2}$, with
\begin{equation}
n_2' \equiv Int[\frac{\gamma B+N_b|c^{ab}|}{c^{a}}-\frac{1}{2}],
\end{equation}
now $0\leq Int(x) \leq N_a$, $S_b^m=N_b$,  $S^m= n_2 + N_b$.

\subsubsection{$c^{a} \geq 2\gamma B-2 N_b c^{ab}$}

$S_a^m = 0$, $S_b^m =S^m = N_b$.


\section{$S_a^m$, $S_b^m$ and $S^m$ for $0<c^{ab}\leq{2\gamma B}$, $c^{a}\geq c^{b}>c^{ab}$, $N_a=N_b=N$ \label{b}}

\begin{figure}
\begin{center}
\scalebox{0.6}{\includegraphics[146,434][482,695]{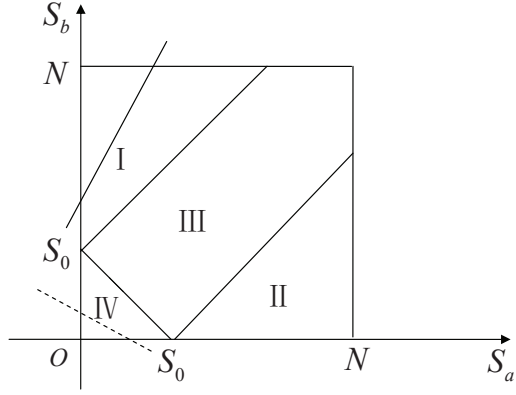}}
\end{center}
\caption{For $0<c^{ab}\leq 2\gamma B$ and $N_a=N_b=N$, the whole region of $(S_a, S_b)$, which satisfy $0 \leq S_a \leq N$ and $0 \leq S_b \leq N$, can be divided into four regions. (I) $S_b-S_a \geq S_0 \geq 0$, (III) $|S_b-S_a|\leq S_0\leq S_b+S_a$, and the rest regions of (II) and (III). The dashed line represents
$\frac{\partial{E}}{\partial{S_b}}=0$ in the case $c^{a} \geq c^{b}>c^{ab}$ and $c^{b}<2\gamma B$. The solid line represents
$\frac{\partial E}{\partial S_b}=0$ in the case $c^{a} > c^{ab} > c^{b}$ and  $c^{a}c^{b} > (c^{ab})^2$. } \label{figb2}
\end{figure}

Define
\begin{equation}
S_0 \equiv \frac{\gamma B}{c^{ab}}-\frac{1}{2},
\end{equation}
which is the value of $S$ on which $S$-dependent part of $E$ is minimal if there were no constraint on $S$.

With $0<c^{ab}\leq{2\gamma B}$, it can be found that the whole region of $(S_a, S_b)$ can be divided into four regions, as shown in FIG.~\ref{figb2}.

In region I,
$S_b-S_a \geq{S_0} \geq 0$, hence $E$ is minimal when $S=S_b-S_a$, with
\begin{widetext}
\begin{equation}
E  = \frac{c^{a}}{2}S_a(S_a+1)+\frac{c^{b}}{2}S_b(S_b+1)-c^{ab}S_aS_b -c^{ab}S_a-\gamma B(S_b-S_a), \label{eb1}
\end{equation}
\end{widetext}
for which it is found that $
\frac{\partial{E}}{\partial{S_b}}>0$.
Thus in region I,  $E$ reaches its minimum at $S_b=S_a+S_0$. It is then easy to note that the minimum
of $E$ in this region rests on $S_a=0$, $S_b=S_0$.

Similarly, it can be shown that in region II
the minimum of $E$ rests on $S_b=0$, $S_a=S_0$. Since both
$(0, S_0)$ and $(S_0,0)$ also belong to region III, the minimum of $E$ in the whole rectangular must be in regions III and IV.

In region III, $|S_b-S_a|\leq{S_0}\leq{S_b+S_a}$, hence $S$ can reach $S_0$, hence
\begin{widetext}
\begin{equation}
E
=\frac{c^{a}-c^{ab}}{2} S_a(S_a+1)+\frac{c^{b}-c^{ab}}{2}S_b(S_b+1)+
\frac{c^{ab}}{2}S_0(S_0+1) - \gamma B S_0, \label{eb3}
\end{equation}
\end{widetext}
for which
$\frac{\partial{E}}{\partial{S_a}} >0$,
$\frac{\partial{E}}{\partial{S_b}}
=\frac{c^{b}-c^{ab}}{2}(2S_b+1)
>0.$
Thus the minimum of $E$ in region III must rest on the border between III and IV, defined by $S_a+S_b=S_0$.

To conclude the above discussion, the minimum $E$ must locate in region IV, where $S_a+S_b \leq S_0$, hence the minimum of $E$ lies on $S=S_a+S_b$,
\begin{widetext}
\begin{equation}
E=\frac{c^{a}}{2}S_a(S_a+1)+\frac{c^{b}}{2}S_b(S_b+1)+c^{ab}S_aS_b
-\gamma B(S_a+S_b). \label{e3}
\end{equation}
\end{widetext}
One obtains
\begin{equation}
\frac{\partial{E}}{\partial{S_a}}=c^{a}S_a+c^{ab}S_b+\frac{c^{a}}{2}-\gamma B,
\end{equation}
\begin{equation}
\frac{\partial{E}}{\partial{S_b}}=c^{b}S_b+c^{ab}S_a+\frac{c^{b}}{2}-\gamma B,
\end{equation}
according to which one needs to consider two subcases.

\subsection{$c^{b}\geq {2\gamma B}$}

In this parameter regime,  $\frac{\partial{E}}{\partial{S_a}}\geq{0}$,
$\frac{\partial{E}}{\partial{S_b}}\geq{0}$. Therefore $E$ is minimal when $S^m_a=S^m_b=S^m=0$.

\subsection{$c^{b}<2\gamma B$}

In this parameter regime,  $\frac{\partial{E}}{\partial{S_b}}=0$
defines a stationary line, shown as the dashed line in FIG.~\ref{figb2}. Consequently, the minima of $E$ in different parts of region IV are
\begin{widetext}
\begin{equation}
 E=\left\{ \begin{array}{ll}
 \frac{c^{a}c^{b}-(c^{ab})^2}{2c^{b}}S_a^2+
 [\frac{c^{a}-c^{ab}}{2}
 -\frac{c^{b}-c^{ab}}{c^{b}}\gamma B]S_a
 -\frac{c^{b}}{8} - \frac{\gamma^2 B^2}{2c^{b}},
 & \text{if} \quad 0\leq{S_a}<\frac{\gamma B}{c^{ab}}-\frac{c^{b}}{2c^{ab}}, \label{x1} \\
 \frac{c^{a}}{2}S_a(S_a+1)-\gamma BS_a, &  \text{if} \quad \frac{\gamma B}{c^{ab}}
 -\frac{c^{b}}{2c^{ab}}\leq S_a\leq{S_0},
 \end{array} \right.
 \end{equation}
 \end{widetext}
For $S_a\geq \frac{\gamma B}{c^{ab}}-\frac{c^{b}}{2c^{ab}}$, it is found that $\frac{\partial{E}}{\partial{S_a}}>0$, hence  $E$ reaches
its minimum at $(\frac{\gamma B}{c^{ab}}-\frac{c^{b}}{2c^{ab}},0)$, i.e. the point bordering the other part of region IV.

Therefore, the minimum
of $E$ in the whole rectangular must locate on the dashed line in sector IV, on which $E$ is given by (\ref{x1}).
Then there are two possibilities.

\subsubsection{$c^{b}<2\gamma B\leq{\frac{c^{a}-c^{ab}}{c^{b}-c^{ab}}}c^{b}$}

We have $S^m_a=0$, $S^m_b=S^m=\frac{\gamma B}{c^{b}}-\frac{1}{2}$.

\subsubsection{$2\gamma B>\frac{c^{a}-c^{ab}}{c^{b}-c^{ab}}c^{b}$}

One finds that
\begin{eqnarray}
S^m_a & = & Int\left[\frac{\gamma B(c^{b}-c^{ab})-c^{b}(c^{a}-c^{ab})/2)}
{c^{a}c^{b}-(c^{ab})^2}\right],\\
S^m_b & = & \frac{\gamma B-c^{ab}S_a}{c^{b}}-\frac{1}{2}, \\
S^m & =&  S^m_a+S^m_b. \\
\end{eqnarray}

\section{$S_a^m$, $S_b^m$ and $S^m$ for $0<c^{ab}\leq{2\gamma B}$, $c^{a} > {c^{ab}} > {c^{b}}$,    $c^{a}c^{b}>{(c^{ab})^2}$, $N_a=N_b=N$ \label{c}}

Again, we use FIG.~\ref{figb2}. It can be shown that in region IV,
$\frac{\partial{E}}{\partial{S_b}} < 0$,
thus the minimum of $E$ in this region lies on the border line  with region III, i.e. $S_a+S_b=S_0$. It can be shown that the
minimum of $E$ in region II lies on the border line $S_a-S_b=S_0$ with region III. Therefore we need only to consider regions I and III.

In region III, as shown in last section, $S^m = S_0$, and $E$ is  given in Eq.~(\ref{e3}). But now that $c^{b}\leq{c^{ab}}\leq{c^{a}}$, we have $\frac{\partial{E}}{\partial{S_a}}>0$
and $\frac{\partial{E}}{\partial{S_b}}<0$. Consequently the minimum of $E$ lies in the border line $S_b-S_a=S_0$.

Therefore, we conclude that $E$ takes its global minimum in region I, where, as discussed in last section, $S^m=S_b-S_a$, $E$ is as given in Eq.~(\ref{eb1}), for which
\begin{equation}
\frac{\partial{E}}{\partial{S_b}} = c^{b}S_b-c^{ab}S_a+\frac{c^{b}}{2}-\gamma B.
\end{equation}
As shown in FIG.~\ref{figb2}, $\frac{\partial{E}}{\partial{S_b}}=0$ defines a  stationary line which crosses with $S_b=N$ at
$(\frac{c^{b}N+c^{b}/2-\gamma B}{c^{ab}}, N)$. The minima of $E$ are found to be:
 \begin{widetext}
 \begin{equation}
 E=\left\{ \begin{array}{ll}
 \frac{c^{a}c^{b}-(c^{ab})^2}{2c^{b}}S_a^2+[\frac{c^{a}-c^{ab}}{2}
 -\frac{c^{ab}-c^{b}}{c^{b}}\gamma B]S_a+ (\frac{3\gamma B}{2} + \frac{c^b}{4})(\frac{\gamma B}{c^b}-\frac{1}{2}) ,
 & \text{if} \quad 0\leq{S_a}<\frac{c^{b}N+c^{b}/2-\gamma B}{c^{ab}}, \\
 \frac{c^{a}}{2}S_a(S_a+1)-c^{ab}(N+1)S_a+\gamma B S_a+
 \frac{c^{b}}{2} N(N+1)-\gamma B N, & \text{if} \quad \frac{c^{b}N+c^{b}/2-\gamma B}{c^{ab}}
 \leq S_a \leq {N-S_0}.
 \end{array} \right.
 \end{equation}
 \end{widetext}
 In the second interval $\frac{c^{b}N+c^{b}/2-\gamma B}{c^{ab}}  \leq S_a\leq{N-S_0}$,
 \begin{eqnarray}
 \frac{\partial{E}}{\partial{S_a}}
 &=&c^{a}S_a-c^{ab}(N+1)+\gamma B+\frac{c^{a}}{2} \nonumber  \\
 & \geq & \frac{c^{a}c^{b}-(c^{ab})^2}{c^{ab}}N
 +\frac{c^{a}}{2}(\frac{c^{b}}{c^{ab}}+1) \nonumber \\
 & & -c^{ab}-\gamma B(\frac{c^{a}}{c^{ab}}-1), \label{derivative2}
 \end{eqnarray}
which is positive if $N> N^*$, where
\begin{equation}
N^* \equiv \frac{\gamma B (c^{a}-c^{ab})+ (c^{ab})^2 -
c^{ab} c^{a} -\frac{c^{a}c^{b}}{2}}{c^{a}c^{b}-(c^{ab})^2}.
\label{nn}
\end{equation}
Then the minimum of $E$ must locate on the stationary line
 $c^{b}S_b-c^{ab}S_a+\frac{c^{b}}{2}-\gamma B=0$, with $0\leq{S_a}<\frac{c^{b}N+c^{b}/2-\gamma B}{c^{ab}}$. One can see two possibilities.

\subsection{$2\gamma B\leq \frac{c^{a}-c^{ab}}{c^{ab}-c^{b}}c^{b}$}

In this case, $\frac{c^{a}-c^{ab}}{2}
 -\frac{c^{ab}-c^{b}}{c^{b}}\gamma B\geq 0$. Thus we have $S_a^m=0$, $S_b^m=S^m=Int[\frac{\gamma B}{c^{b}}-\frac{1}{2}]$.

\subsection{$2\gamma B> \frac{c^{a}-c^{ab}}{c^{ab}-c^{b}}c^{b}$}

Then
\begin{eqnarray}
S_a^m & = &Int[\frac{\gamma B(c^{ab}-c^{b})-c^{b}(c^{a}-c^{ab})/2)}
{c^{a}c^{b}-(c^{ab})^2}], \\
S_b^m & = & Int[\frac{\gamma B+c^{ab}S^m_a}{c^{b}}-\frac{1}{2}], \\
S^m & = & S^m_b-S^m_a.
\end{eqnarray}

\section{$S_a^m$, $S_b^m$ and $S^m$ for $0<c^{ab}\leq{2\gamma B}$, $c^{b}\geq c^{a}>c^{ab}$, $N_a=N_b=N$ \label{bb}}

By exchanging the labels $a$ and $b$ in Appendix~\ref{b}, one obtains the following results.

\subsection{$c^{a}\geq {2\gamma B}$}

$S^m_a=S^m_b=S^m=0$.

\subsection{$c^{a}<2\gamma B$}

\subsubsection{$c^{a}<2\gamma B\leq{\frac{c^{b}-c^{ab}}{c^{a}-c^{ab}}}c^{a}$}

We have $S^m_b=0$, $S^m_a=S^m=\frac{\gamma B}{c^{a}}-\frac{1}{2}$.

\subsubsection{$2\gamma B>\frac{c^{b}-c^{ab}}{c^{a}-c^{ab}}c^{a}$}

\begin{eqnarray}
S^m_b & = & Int\left[\frac{\gamma B(c^{a}-c^{ab})-c^{a}(c^{b}-c^{ab})/2)}
{c^{a}c^{b}-(c^{ab})^2}\right],\\
S^m_a & = & \frac{\gamma B-c^{ab}S_b}{c^{a}}-\frac{1}{2}, \\
S^m & =&  S^m_a+S^m_b. \\
\end{eqnarray}

\section{$S_a^m$, $S_b^m$ and $S^m$ for $0<c^{ab}\leq{2\gamma B}$, $c^{b} > {c^{ab}} > {c^{a}}$,    $c^{a}c^{b}>{(c^{ab})^2}$, $N_a=N_b$ }

By exchanging the labels $a$ and $b$ in Appendix~\ref{c}, one obtains the following results.

\subsection{$2\gamma B\leq \frac{c^{b}-c^{ab}}{c^{ab}-c^{a}}c^{a}$}

$S_b^m=0$, $S_a^m=S^m=Int[\frac{\gamma B}{c^{a}}-\frac{1}{2}]$.

\subsection{$2\gamma B> \frac{c^{b}-c^{ab}}{c^{ab}-c^{a}}c^{a}$}

\begin{eqnarray}
S_b^m & = &Int[\frac{\gamma B(c^{ab}-c^{a})-c^{a}(c^{b}-c^{ab})/2)}
{c^{a}c^{b}-(c^{ab})^2}], \\
S_a^m & = & Int[\frac{\gamma B+c^{ab}S^m_b}{c^{a}}-\frac{1}{2}], \\
S^m & = & S^m_b-S^m_a.
\end{eqnarray}

\end{document}